\documentclass[10pt,twocolumn]{IEEEtran}

\usepackage[latin1]{inputenc}
\usepackage{times}
\usepackage{calc}
\usepackage[dvips]{color}
\usepackage{cite}
\usepackage{graphicx}
\usepackage{amsmath}
\usepackage{amssymb}
\usepackage{caption}
\usepackage{subfigure}
\usepackage{colortbl}
\graphicspath{{eps/}}

\usepackage{bbm}
\usepackage{ifthen}

\newcommand{\comment}[1]{}

\newtheorem{lemma}{Lemma}
\newtheorem{lp}{{\bf LP}}
\newtheorem{conjecture}{Conjecture}
\newtheorem{example}{Example}

\newcommand{\btheorem}{\begin{theorem}\rm}
\newcommand{\etheorem}{\end{theorem}}
\newcommand{\blemma}{\begin{lemma}\rm}
\newcommand{\blemmaalphaformula}{\begin{lemmaalphaformula}\rm}
\newcommand{\elemma}{\end{lemma}}
\newcommand{\bconjecture}{\begin{conjecture}\rm}
\newcommand{\econjecture}{\end{conjecture}}
\newcommand{\bexample}{\begin{example}\rm}
\newcommand{\eexample}{\end{example}}

\newcommand{\nldpc}[3]{\text{LDPC}(#1, #2, #3)}
\newcommand{\eldpc}[3]{\text{LDPC}(#1, #2, #3)}

\newcommand{\graph}{{{\tt G}}}

\newcommand{\ssetsize}{s}
\newcommand{\ssetsizemin}{s_{\text{min}}}

\newcommand{\edgesetsize}{e}

\newcommand{\ssmin}{s_{\text{min}}}

\newcommand{\defeq}{\triangleq}

\newcommand{\ind}{\mathbf{1}}
\newcommand{\variance}{{\mathcal V}}

\newcommand{\Pt}{{\prob_{\text{target}}}}
\newcommand{\Vtree}{{{\tt V}}}
\newcommand{\Ctree}{{{\tt C}}}
\newcommand{\tree}{{\tt T}}
\newcommand{\btree}{{\tt T^c}}
\newcommand{\grapht}{{{\tt G_{T}}}}

\newcommand{\Vgtp}{V^{\grapht'}}

\newcommand{\ledge}{\lambda}    
\newcommand{\redge}{\rho}       
\newcommand{\lnode}{\Lambda}    
\newcommand{\lnodenormalized}{L}    
\newcommand{\rnodenormalized}{R}    

\newcommand{\coef}[2]{\text{coef}\left\{ #1, #2\right\}}

\newcommand{\expectation}{\ensuremath{\mathbb{E}}}

\newcommand{\dlmax}{{\tt l}_{\text{max}}} 
\newcommand{\dr}{{\tt r}} 
\newcommand{\drmax}{{\tt r}_{\text{max}}} 
\newcommand{\dl}{{\tt l}}

\DeclareMathOperator{\prob}{{\mbox{\rm P}}}

\begin{document}

\title{How to Find Good Finite-Length Codes:\\ From Art Towards Science}
\author{
\thanks{
This invited paper is an enhanced version of the work presented at the $4^{\rm th}$ International Symposium on Turbo Codes and Related Topics, M{\"u}nich, Germany, 2006.}
\thanks{The work presented in this paper was supported (in part) by the National Competence Center 
in Research on Mobile Information and Communication Systems (NCCR-MICS), a center 
supported by the Swiss National Science Foundation under grant number 5005-67322.}
\thanks{AM has been partially supported by the EU under the integrated project
EVERGROW.}
Abdelaziz Amraoui$^{\dagger}$ 
\thanks{$^{\dagger}$ A.~Amraoui is with EPFL, School of Computer and Communication Sciences, 
Lausanne CH-1015, Switzerland, {\tt abdelaziz.amraoui@epfl.ch} }, 
Andrea Montanari$^*$ 
\thanks{$^*$ A.~Montanari is with Ecole Normale Sup\'{e}rieure, 
Laboratoire de Physique Th\'{e}orique, 75231 Paris Cedex 05, France, 
{\tt montanar@lpt.ens.fr}}
and Ruediger Urbanke$^{\ddagger}$ 
\thanks{$^{\ddagger}$ R.~Urbanke is with EPFL, School of Computer and Communication Sciences, 
Lausanne CH-1015, Switzerland, {\tt ruediger.urbanke@epfl.ch}}}

\maketitle


\begin{abstract}
We explain how to optimize finite-length LDPC codes for transmission over the binary erasure
channel. Our approach relies on an analytic approximation of the erasure probability.
This is in turn based on a finite-length 
scaling result to model large scale erasures and a union bound involving minimal stopping sets
to take into account small error events.
We show that the performances of optimized ensembles as 
observed in simulations are well described by our approximation. 
Although we only address the case of transmission over the binary erasure channel,
our method should be applicable to a more general setting.
\end{abstract}

%
%

\section{Introduction}\label{sec:intro}
%
In this paper, we consider transmission using random elements from the standard
ensemble of low-density parity-check (LDPC) codes defined by the 
degree distribution pair $(\ledge, \redge)$.
For an introduction to LDPC codes and the standard notation see \cite{RiU05}.
In \cite{Mon01b}, one of the authors (AM) suggested that the probability
of error of iterative coding systems follows a scaling law. 
In \cite{AMRU04,AMRU04-allerton,AMU05}, it was shown that this
is indeed true for LDPC codes, assuming that transmission takes
place over the BEC.
Strictly speaking, scaling laws describe the asymptotic behavior
of the error probability close to the threshold for increasing blocklengths.
However, as observed empirically in the papers  mentioned  above,
scaling laws provide good approximations to the error probability  also away
from the threshold and already for modest blocklengths. 
This is the starting point for our finite-length optimization. 

In \cite{AMRU04,AMU05} the {\em form} of the scaling law for transmission
over the BEC was derived and it was shown how to compute the {\em scaling parameters} 
by solving a system of ordinary differential equations. This system was called
{\em covariance evolution} in analogy to density evolution. 
Density evolution concerns the evolution of the {\em average}
number of erasures still contained in the graph during the decoding process, 
whereas covariance evolution concerns the evolution
of its {\em variance}. Whereas Luby et al. \cite{LMSSS97} found
an explicit solution to the density evolution equations, 
to date no such solution is known for the system of covariance
equations. Covariance evolution must therefore be integrated numerically.
Unfortunately the dimension of the ODE's system
ranges from hundreds to thousand for typical examples.
As a consequence, numerical integration can be quite
time consuming. This is a serious problem if we want to 
use scaling laws in the context of optimization, where the 
computation of scaling parameters must be repeated for many different 
ensembles during the optimization process.

In this paper, we make two main contributions. First, we derive 
explicit analytic expressions for the scaling parameters as a function 
of the degree distribution pair and quantities which appear in density 
evolution. Second, we provide  
an accurate approximation to the erasure probability 
stemming from small stopping sets and resulting in the erasure floor. 

The paper is organized as follows. Section \ref{sec:scaling}
describes our approximation for the error probability, the
scaling law being discussed in Section
\ref{sec:waterfall} and the error floor in Section~\ref{sec:errorfloor}.
 We combine
these results and give in Section~\ref{sec:approximation} an  
approximation to the erasure probability curve,
denoted by $\prob(n, \ledge, \redge, \epsilon)$,
that can be computed efficiently for any 
blocklength, degree distribution pair, and channel parameter. 
The basic ideas behind the explicit determination of the scaling 
parameters (together with the resulting expressions)
are collected in Section~\ref{sec:alternative}.
Finally, the most technical (and tricky) part of this computation is
deferred to Section \ref{sec:messagevariance}.

As a motivation for some of the rather technical
points to come, we start in Section \ref{sec:optimization} 
by showing how $\prob(n, \ledge, \redge, \epsilon)$
can be used to perform an efficient finite-length optimization. 

\subsection{Optimization}
\label{sec:optimization}
The optimization procedure takes as input a  blocklength $n$, the BEC 
erasure probability $\epsilon$, and a target probability of 
erasure, call it $\Pt$. Both bit or block probability can be considered.
We want to find a degree distribution pair
$(\ledge, \redge)$ of maximum rate so that 
$\prob(n, \ledge, \redge, \epsilon) \leq \Pt$, where
$\prob(n, \ledge, \redge, \epsilon)$ is the approximation discussed in
the introduction.

Let us describe an efficient procedure to accomplish this
optimization locally (however, many equivalent approaches are possible). 
Although providing a global optimization scheme
goes beyond the scope of this paper, the local procedure was found
empirically to converge often to the global optimum.

It is well known \cite{RiU05} that the {\em design rate} 
$r(\ledge,\redge)$ associated to a degree distribution pair $(\ledge, \redge)$ is
equal to
\begin{align}
r(\ledge, \redge) = & 1- \frac{\sum_i \frac{\redge_i}{i}}{\sum_j \frac{\ledge_j}{j}}.
\end{align}
For ``most'' ensembles the actual rate of a randomly chosen element of the ensemble 
$\eldpc n \ledge \redge$ is close to this
design rate \cite{MMU05}. In any case, $r(\ledge, \redge)$ 
is {\em always} a lower bound.
Assume we change the degree distribution pair slightly by
$\Delta\ledge(x)=\sum_i \Delta \ledge_i x^{i-1}$ and 
$\Delta\redge(x)=\sum_i \Delta\redge_i x^{i-1}$,
where $\Delta \ledge(1)=0=\Delta\redge(1)$ and assume that the change is
sufficiently small so that 
$\ledge+\Delta \ledge$ as well as 
$\redge+\Delta \redge$ are still valid degree distributions (non-negative
coefficients). A quick calculation then shows that
the design rate changes by
\begin{align}
& r(\ledge+\Delta\ledge, \redge+\Delta\redge) - r(\ledge, \redge)
\nonumber\\ 
& \simeq \sum_{i} \Delta \ledge_i
\frac{(1-r)}{i\sum_j\frac{\ledge_j}{j}} - \sum_{i}
\frac{\Delta \redge_i}{i\sum_j\frac{\ledge_j}{j}}.\label{equ:rate}
\end{align}
In the same way, the erasure probability changes (according
to the approximation) by
\begin{align}
& \prob(n, \ledge+\Delta\ledge, \redge+\Delta\redge, \epsilon) - 
\prob(n, \ledge,\redge, \epsilon) \nonumber\\
&  \simeq \sum_{i} \Delta \ledge_i \frac{\partial \prob}{\partial \ledge_i} 
+ \sum_{i} \Delta \redge_i \frac{\partial \prob}{\partial \redge_i}.\label{equ:PB}
\end{align}
Equations (\ref{equ:rate}) and (\ref{equ:PB}) give rise to a simple linear
program to optimize locally the degree distribution: Start with some initial
degree distribution pair $(\ledge, \redge)$. 
If $\prob(n, \ledge,\redge, \epsilon) \leq\Pt$, then increase the rate
by a repeated application of the following linear program.

\begin{lp}\rm[Linear program to increase the rate]
\label{lp1}
\begin{align*}
\max\{
& (1-r) \sum_i \Delta \ledge_i/i-\sum_i \Delta \redge_i/i \mid\\
& \sum_i \Delta \ledge_i=0; \quad  \quad \quad-\min\{\delta, \ledge_i\} \leq \Delta \ledge_i \leq \delta ;  \\
& \sum_i \Delta \redge_i=0;  \quad  \quad \quad-\min\{\delta, \redge_i\} \leq \Delta \redge_i \leq \delta ;  \\
& \sum_i \frac{\partial \prob}{\partial \ledge_i} \Delta \ledge_i + 
\sum_i \frac{\partial \prob}{\partial \redge_i} \Delta \redge_i 
\leq \Pt-\prob(n,\ledge, \redge,\epsilon) 
\}.
\end{align*}
\end{lp}
Hereby, $\delta$ is a sufficiently small 
non-negative number to ensure that the degree distribution pair
changes only slightly at each step so that changes of the rate and
of the probability of erasure are accurately described by the linear 
approximation. The value $\delta$ is best adapted dynamically
to ensure convergence.
One can start with a large value and decrease it the closer we
get to the final answer.
The objective function in LP \ref{lp1} is equal to the total derivative
of the rate as a function of the change of the degree distribution. 
Several rounds of this linear program will gradually improve the rate of the 
code ensemble, while keeping the erasure probability below the target
(last inequality).
 
Sometimes it is necessary to initialize the optimization procedure
with degree distribution pairs that do not fulfill
the target erasure probability constraint. 
This is for instance the case if the optimization is repeated for
a large number of ``randomly'' chosen initial conditions.
In this way, we can check whether the procedure always converges to the 
same point (thus suggesting that a global optimum was found), or otherwise, 
pick the best outcome of many trials. To this end we define a linear program
that decreases the erasure probability.
\begin{lp}\rm[Linear program to decrease $\prob(n,\ledge,\redge,\epsilon)$]
\label{lp2}
\begin{align*}
\min\{& \sum \frac{\partial \prob}{\partial \ledge_i} \Delta \ledge_i + 
\sum \frac{\partial \prob}{\partial \redge_i} \Delta \redge_i \mid\\
& \sum_i \Delta \ledge_i=0; \quad  \quad \quad-\min\{\delta, \ledge_i\} \leq \Delta \ledge_i \leq \delta ;  \\
& \sum_i \Delta \redge_i=0;  \quad  \quad \quad-\min\{\delta, \redge_i\} \leq \Delta \redge_i \leq \delta
\}.
\end{align*}
\end{lp}

\bexample[Sample Optimization]
\label{sec:optimexample}
Let us show a sample optimization.
Assume we transmit over a BEC with channel erasure probability $\epsilon=0.5$. 
We are interested in a block length of $n=5000$ bits and the maximum variable 
and check degree we allow are $\dlmax=13$ and $\drmax=10$, respectively.
We constrain the {\em block} erasure probability
to be smaller than $\Pt=10^{-4}$.
We further count only erasures larger or equal to $\ssetsizemin=6$ bits. This
corresponds to looking at an {\em expurgated} ensemble, i.e., we are looking at the
subset of codes of the ensemble that do not contain 
stopping sets of sizes smaller than $6$. Alternatively, we can interpret this
constraint in the sense that we use an outer code which
``cleans up'' the remaining small erasures. Using the techniques
discussed in Section \ref{sec:errorfloor}, we can compute the probability that
a randomly chosen element of an ensemble does not contain stopping
sets of size smaller than $6$. If this probability is not too small
then we have a good chance of finding such a code in the ensemble by
sampling a sufficient number of random elements. 
This can be checked at the end of the optimization procedure.

\begin{figure}[htp]
\centering
\setlength{\unitlength}{0.35bp}%
\begin{picture}(490, 308)
\put(0, 0){\includegraphics[scale=0.35]{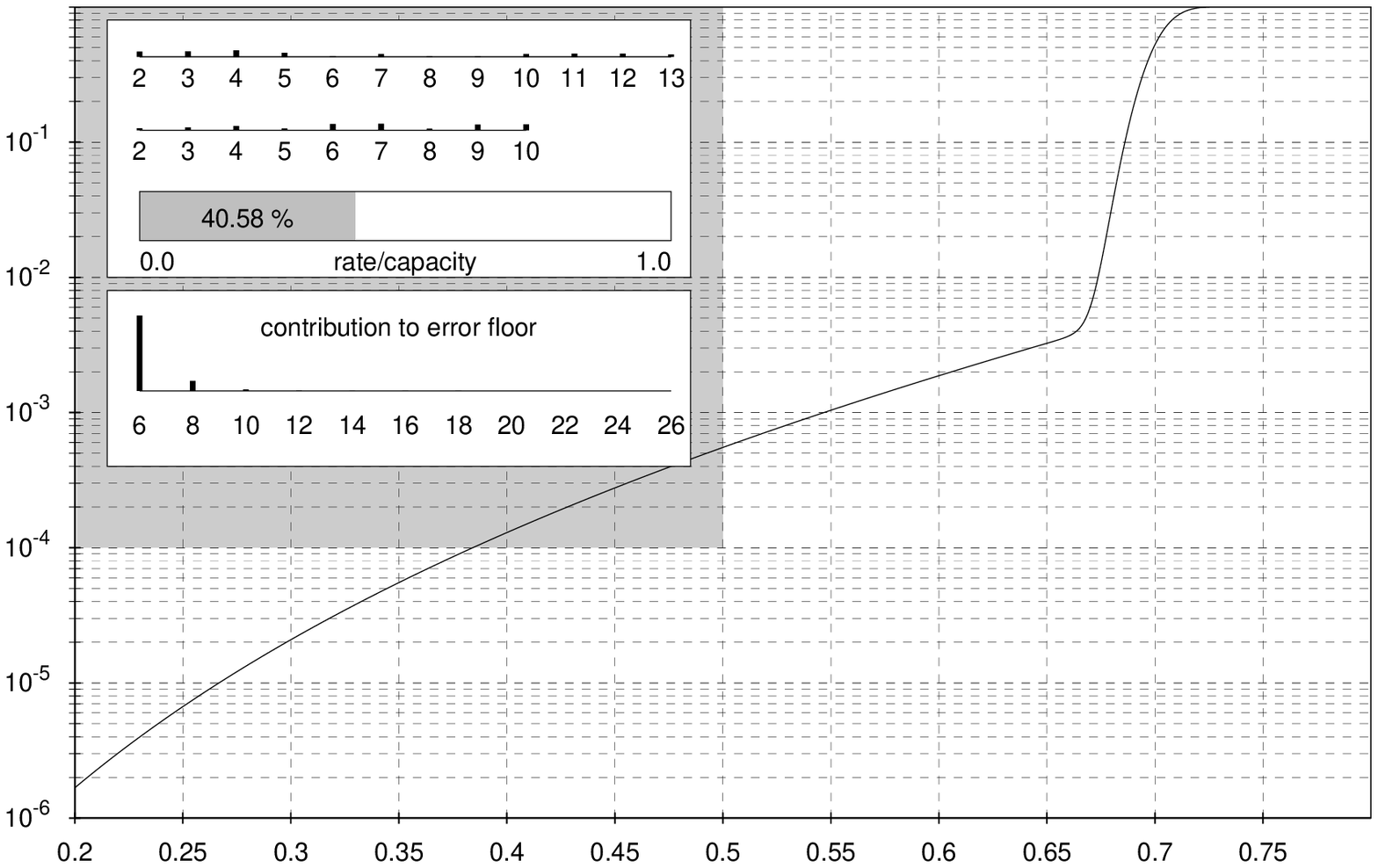}}
\put(480,9){\makebox(0,0){$\epsilon$}}
\put(11,300){\makebox(0,0){$P_B$}}
\end{picture}
\caption{\label{fig:optimstart} 
Approximation of the block erasure probability for 
the initial ensemble with degree distribution pair given in
(\ref{equ:lambda0}) and (\ref{equ:rho0}).}
\end{figure}

We start with an arbitrary degree distribution pair:
\begin{align}
\lambda(x) & = 0.139976 x +0.149265 x^2 +0.174615 x^3 \label{equ:lambda0}\\
& +0.110137 x^4 +0.0184844 x^5 + 0.0775212 x^6\nonumber\\
& + 0.0166585 x^7 +0.00832646 x^8 + 0.0760256 x^9\nonumber\\
& + 0.0838369 x^{10} + 0.0833654 x^{11} + 0.0617885 x^{12},\nonumber\\
\redge(x)=&0.0532687 x +0.0749403 x^2 +0.11504 x^3\label{equ:rho0}\\
&  +0.0511266 x^4 +0.170892 x^5 +0.17678 x^6\nonumber\\
& +0.0444454 x^7  +0.152618 x^8 + 0.160889 x^9.\nonumber
\end{align}
This pair was generated randomly by choosing 
each coefficient uniformly in
$[0, 1]$ and then normalizing so that $\ledge(1)=\redge(1)=1$.
The approximation of the block erasure probability 
curve of this code (as given in Section \ref{sec:approximation}) is shown in
Fig.~\ref{fig:optimstart}.
For this initial degree distribution pair we have 
$r(\ledge,\redge)=0.2029$ and 
$\prob_B(n=5000,\ledge,\redge,\epsilon=0.5)=0.000552>\Pt$.
Therefore, we start by reducing 
$\prob_B(n=5000, \ledge,\redge, \epsilon=0.5)$ (over the choice of $\ledge$ and
$\redge$) using LP~\ref{lp2} until it becomes lower than $\Pt$. 
\begin{figure}[htp]
\centering
\setlength{\unitlength}{0.35bp}%
\begin{picture}(490, 308)
\put(0, 0){\includegraphics[scale=0.35]{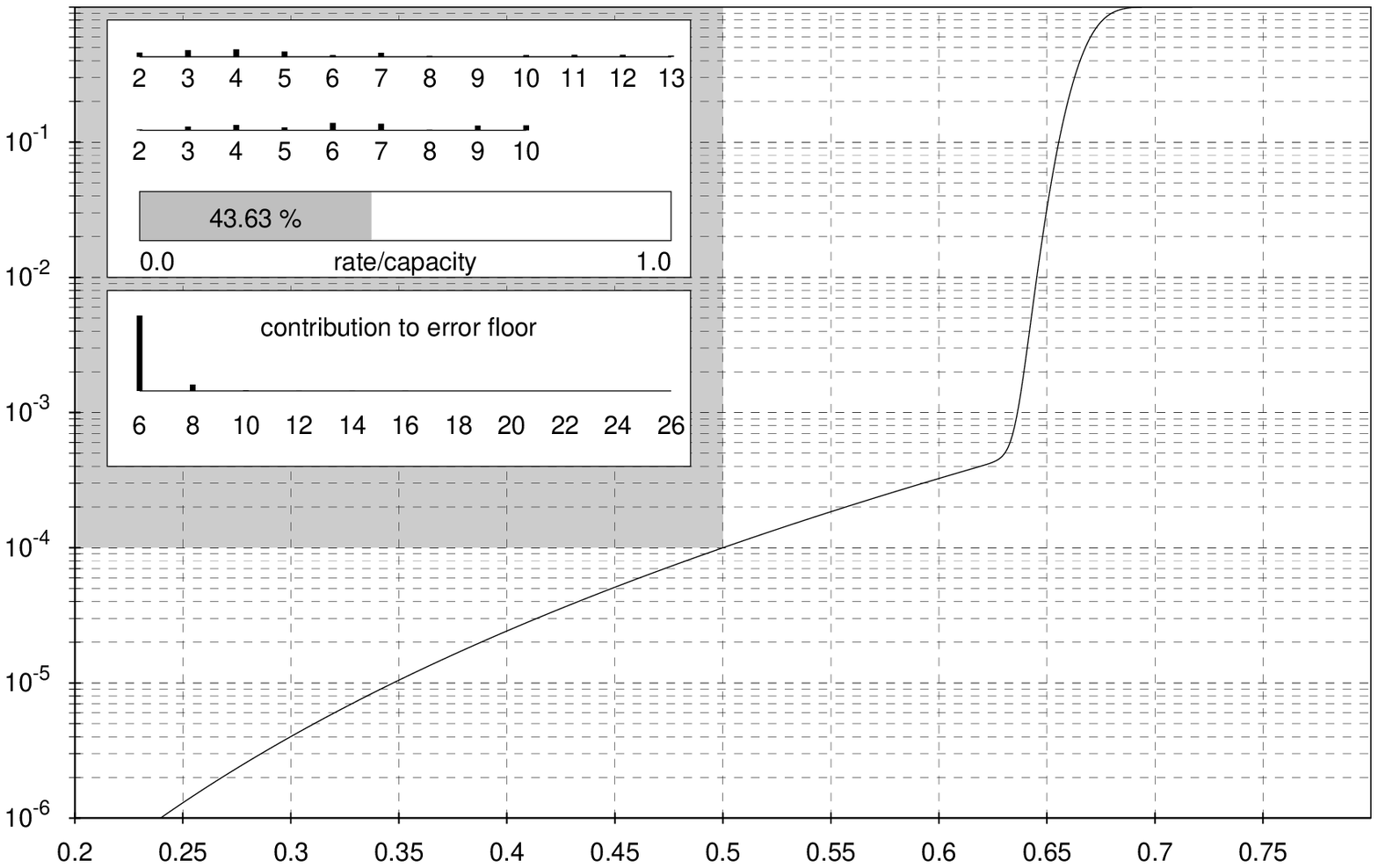}}
\put(480,9){\makebox(0,0){$\epsilon$}}
\put(11,300){\makebox(0,0){$P_B$}}
\end{picture}
\caption{\label{fig:optimmiddle} Approximation of the block erasure probability for the ensemble obtained after the first part of the optimization (see (\ref{equ:lambda1}) and (\ref{equ:rho1})). The erasure probability has been lowered below the target.}
\end{figure}
After a number of LP~\ref{lp2} rounds we obtain the degree distribution pair:
\begin{align}
\ledge(x)=& 
0.111913 x +
0.178291 x^2 +
0.203641 x^3 \label{equ:lambda1}\\& +
0.139163 x^4 +
0.0475105 x^5 +
0.106547 x^6 \nonumber\\ &+
0.0240221 x^7 +
0.0469994 x^9 +
0.0548108 x^{10} \nonumber\\ &+
0.0543393 x^{11} +
0.0327624 x^{12},\nonumber\\
\redge(x)=& 
0.0242426 x +
0.101914 x^2 +
0.142014 x^3 \label{equ:rho1}\\ &+
0.0781005 x^4 +
0.198892 x^5 +
0.177806 x^6  \nonumber\\ &+
0.0174716 x^7 +
0.125644 x^8 +
 0.133916 x^9.\nonumber
\end{align}
For this degree distribution pair we have
$\prob_B(n=5000,\ledge,\redge, \epsilon=0.5)=0.0000997 \leq \Pt$ and  
$r(\ledge,\redge)=0.218$. We show the corresponding approximation in 
Fig.~\ref{fig:optimmiddle}.

\begin{figure}[htp]
\centering
\setlength{\unitlength}{0.35bp}%
\begin{picture}(490, 308)
\put(0,0){\includegraphics[scale=0.35]{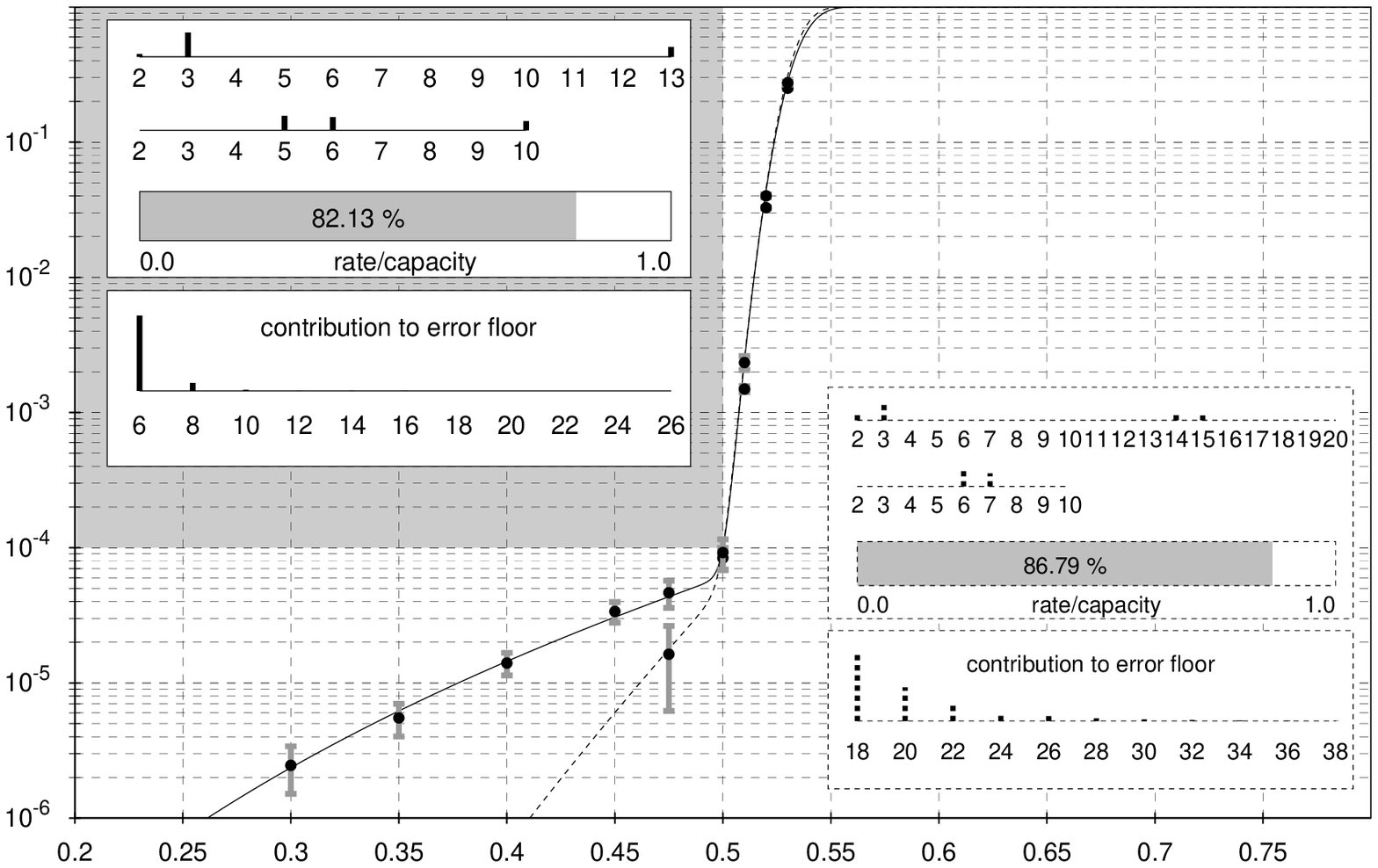}}
\put(480,9){\makebox(0,0){$\epsilon$}}
\put(11,300){\makebox(0,0){$P_B$}}
\end{picture}
\caption{\label{fig:optimfinal} Error probability curve for the result of the
optimization (see (\ref{equ:lambdafinal}) and (\ref{equ:rhofinal})). 
The solid curve is $\prob_B(n=5000, \ledge, \redge,\epsilon=0.5)$ 
while the small dots correspond to simulation points. In dotted are the results 
with a more aggressive expurgation.}
\end{figure}

Now, we start the second phase of the optimization and optimize the rate
while insuring that the block erasure probability remains below the target,
using LP~\ref{lp1}.
The resulting degree distribution  pair is:
\begin{align}
\ledge(x)=& 
0.0739196 x + 0.657891 x^2+ 0.268189 x^{12}\label{equ:lambdafinal},\\
\redge(x)=& 0.390753 x^4+ 0.361589 x^5+ 0.247658 x^9,\label{equ:rhofinal}
\end{align}
where $r(\ledge, \redge)=0.41065$.
The block erasure probability plot for the result of the optimization is
shown in Fig~\ref{fig:optimfinal}. 


Each LP step takes on the order of seconds on a standard PC.
In total, the optimization for a given
set of parameters $(n, \epsilon, \Pt, \dlmax, \drmax, \ssmin)$ takes on the order of  minutes.

Recall that the whole optimization procedure was based on 
$\prob_B(n,\ledge,\redge,\epsilon)$ which is only an
{\em approximation} of the {\em true} block erasure probability.
In principle, the actual performances of the optimized ensemble could
be worse (or better) than predicted by $\prob_B(n,\ledge,\redge,\epsilon)$.
To validate the procedure we computed the block erasure probability
for the optimized degree distribution also by means of simulations 
and compare the two. The simulation results are shown
in Fig~\ref{fig:optimfinal} (dots with $95\%$ confidence intervals) :
analytical (approximate) and numerical 
results are in almost perfect agreement! 

How hard is it to find a code without stopping sets of size smaller than 6
within the ensemble 
$\eldpc {5000} \ledge \redge$ with $(\ledge,\redge)$ given by 
Eqs.~(\ref{equ:lambdafinal}) and (\ref{equ:rhofinal})? 
As discussed in more detail in Section 
\ref{sec:errorfloor}, in the limit of large blocklengths the number
of small stopping sets has a joint Poisson distribution.
As a consequence, if  $\tilde{A}_i$ denotes the expected number 
of minimal stopping sets of size $i$ in a random element from 
$\eldpc {5000} \ledge \redge$, the probability that it contains {\em no}
stopping set of size smaller than $6$ is
approximately $\exp\{-\sum_{i=1}^{5} \tilde{A}_i\}$.
For the optimized ensemble we get $\exp\{-(0.2073+0.04688+0.01676+0.007874+0.0043335)\} \approx 0.753$, a quite large probability.
We repeated the optimization procedure
with various different random initial conditions and always ended up with
essentially the same degree distribution. Therefore, we can be quite confident
that the result of our local optimization is close to the global optimal 
degree distribution pair for the given constraints
$(n, \epsilon, \Pt, \dlmax, \drmax, \ssmin)$.

There are many ways of improving the result. E.g., if we allow higher degrees or 
apply a more aggressive expurgation, we can obtain
degree distribution pairs with higher rate. E.g., for the choice $\dlmax=15$ and $\ssmin=18$ the 
resulting degree distribution pair is
\begin{align}
\ledge(x)=&0.205031 x+ 0.455716 x^2
\label{equ:lambdafinalexpurg},\\
& +0.193248 x^{13}+0.146004 x^{14} \nonumber\\
\redge(x)=& 0.608291 x^5 + 0.391709 x^6, \label{equ:rhofinalexpurg}
\end{align}
where $r(\ledge, \redge)=0.433942$.
The corresponding curve is depicted in 
Fig~\ref{fig:optimfinal}, as a dotted line. 
However, this time the probability that a random element from 
$\eldpc {5000} \ledge \redge$ has {\em no} stopping set of size smaller 
than $18$ is approximately $6.10^{-6}$. It will therefore be
harder to find a code that fulfills the expurgation requirement.
\end{example}
It is worth stressing that our results could be improved further
by applying the same approach to more powerful ensembles, 
e.g., multi-edge type ensembles, or ensembles defined by protographs. 
The steps to be accomplished are: $(i)$
derive the scaling laws and define scaling parameters for such
ensembles; $(ii)$ find efficiently computable expressions for the 
scaling parameters; $(iii)$ optimize the ensemble with respect to its
defining parameters (e.g. the degree distribution) as above.
Each of these steps is a manageable task -- albeit not a trivial one. 

Another generalization of our approach which is slated for future work
is the extension to general binary memoryless symmetric channels. 
Empirical evidence suggests that scaling laws should also hold in this
case, see \cite{Mon01b,AMRU04}. How to prove this fact or how to
compute the required parameters, however, is an open issue.

In the rest of this paper, we describe in detail the approximation
$\prob(n, \ledge, \redge, \epsilon)$ for the BEC.

\section{Approximation $\prob_B(n, \ledge, \redge, \epsilon)$ and $\prob_b(n, \ledge, \redge, \epsilon)$}
\label{sec:scaling}
In order to derive approximations for
the erasure probability we separate the contributions to this erasure probability into
two parts -- the contributions due to large erasure events and the ones
due to small erasure events. The large erasure events give rise to the so-called
{\em waterfall} curve, whereas the small erasure events are responsible for
the {\em erasure floor}.

In Section~\ref{sec:waterfall}, 
we recall that the water fall curve follows a scaling law and we discuss
how to compute the scaling parameters.
We denote this approximation of the water fall curve by 
$\prob_{B/b}^W(n, \ledge, \redge, \epsilon)$.
We next show in Section~\ref{sec:errorfloor} how to approximate
the erasure floor. We call this approximation 
$\prob_{B/b,\ssetsizemin}^E(n, \ledge, \redge, \epsilon)$. Hereby,
$\ssetsizemin$ denotes the {\em expurgation} parameter, i.e, we only count
error events involving at least $\ssetsizemin$ erasures.
Finally, we collect in Section~\ref{sec:approximation} our results
and give an approximation to the total erasure probability.
We start in Section \ref{sec:densityevol} with a short review of
density evolution.

\subsection{Density Evolution}
\label{sec:densityevol}
The initial analysis of the performance of LDPC codes assuming that
transmission takes place of the BEC is due to 
Luby, Mitzenmacher, Shokrollahi, Spielman and Stemann,
see  \cite{LMSSS97}, and it is based on the so-called {\em peeling} algorithm.
In this algorithm we ``peel-off'' one variable node at a time (and all its adjacent check nodes
and edges) creating a sequence of residual graphs. 
Decoding is successful if and only if the final
residual graph is the empty graph. 
A variable node can be peeled off if it is connected to at least
one check node which has residual degree one. 
Initially, we start with the complete Tanner graph representing
the code and in the first step we delete all variable nodes from the 
graph which have been received
(have not been erased), all connected check nodes, and all connected edges.

From the description of the algorithm it should be clear that the number of degree-one check
nodes plays a crucial role. The algorithm stops if and only if no degree-one check node
remains in the residual graph. Luby et al. were able to give analytic expressions for
the expected number of degree-one check nodes as a function of the size of the residual graph
in the limit of large blocklengths. They further showed that most instances of the graph and the
channel follow closely this ensemble average.
More precisely, let $r_1$ denote the fraction 
of degree-one check 
nodes in the decoder. (This means that the actual number of degree-one check nodes
is equal to $n (1-r) r_1$, where $n$ is the blocklength and $r$ is the design rate of
the code.) Then, as shown in \cite{LMSSS97}, $r_1$ is given 
parametrically by
\begin{align}
r_1(y) & =  \epsilon \ledge(y) [y-1+\redge(1-\epsilon \ledge(y))].
\end{align}
where $y$ is determined so that 
$\epsilon \lnodenormalized(y)$ is  the fractional (with respect to $n$) size
of the residual graph. Hereby,
$\lnodenormalized(x)=\sum_i\lnodenormalized_i x^i=
\frac{\int_0^x \ledge(u)du}{\int_0^1 \ledge(u)du}$ is the node perspective
variable node distribution, i.e. $\lnodenormalized_i$ is the fraction of 
variable nodes of degree $i$ in the Tanner graph. Analogously,
we let $\rnodenormalized_i$ denote the fraction of degree $i$
check nodes, and set $\rnodenormalized(x)=\sum_i\rnodenormalized_i x^i$.
With an abuse of notation we shall sometimes denote the irregular 
LDPC ensemble as $\eldpc n \lnodenormalized \rnodenormalized$.

\begin{figure}[htp]
\centering
\setlength{\unitlength}{0.466bp}%
\begin{picture}(367.5, 231)
\put(0,0){\includegraphics[scale=0.466]{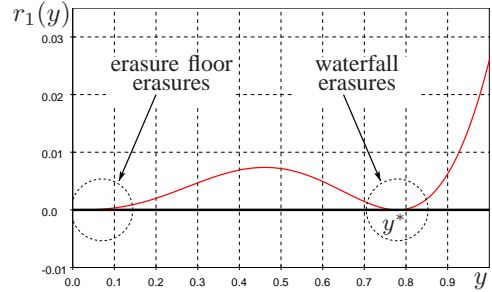}}
\put(360,9){\makebox(0,0){$y$}}
\put(5,225){\makebox(0,0){$r_1(y)$}}
\put(110,185){\makebox(0,0){\small erasure floor}}
\put(110,168){\makebox(0,0){\small erasures}}
\put(260,185){\makebox(0,0){\small waterfall}}
\put(260,168){\makebox(0,0){\small erasures}}
\put(290,55){\makebox(0,0){\small $y^*$}}
\end{picture}
\caption{\label{fig:densityevol} $r_1(y)$ for $y\in[0,1]$ at the threshold.
The degree distribution pair is  $\ledge(x)=x^2$ and $\redge(y)=x^5$ and the 
threshold is $\epsilon^*=0.4294381$.}
\end{figure}

The threshold noise parameter $\epsilon^*=\epsilon^*(\ledge,\redge)$ is the 
supremum value of $\epsilon$ such that $r_1(y)>0$ for all $y\in (0,1]$,
(and therefore iterative decoding is successful with high probability).
In Fig.~\ref{fig:densityevol}, we show the function $r_1(y)$ depicted for the 
ensemble with $\ledge(x)=x^2$ and $\redge(x)=x^5$ for $\epsilon=\epsilon^*$.
As the fraction of degree-one check nodes concentrates around $r_1(y)$, 
the decoder will fail with high probability only in two possible ways.
The first relates to $y \approx 0$ and corresponds to small erasure events. 
The second one corresponds to the value $y^*$ such that $r_1(y^*)=0$.
In this case the fraction of variable nodes that can 
not be decoded concentrates around $\nu^*=\epsilon^* \lnodenormalized(y^*)$.

We call a point $y^*$ where the function $y-1+\redge(1-\epsilon \ledge(y))$
and its derivative both vanish a {\em critical point}.
At threshold, i.e. for $\epsilon = \epsilon^*$, there is at least one critical 
point, but there may be more than one.
(Notice that the function $r_1(y)$ always vanishes together with its derivative
at $y=0$, cf. Fig.~\ref{fig:densityevol}. However, this does not imply that 
$y=0$ is a critical point because of the extra
factor $\ledge(y)$ in the definition of $r_1(y)$.)
Note that if an ensemble has a single critical point and 
this point is strictly positive, then the number of 
remaining erasures conditioned on decoding failure, concentrates
around $\nu^* \defeq \epsilon^* \lnodenormalized(y^*)$.

In the rest of this paper, we will consider ensembles with 
a single critical point and separate the two above contributions. 
We will consider in  Section~\ref{sec:waterfall} erasures of size at 
least $n \gamma \nu^*$ with $\gamma\in (0,1)$.
In Section~\ref{sec:errorfloor} we will instead focus on erasures of size 
smaller than $n\gamma\nu^*$. We will finally combine the two results in 
Section~\ref{sec:approximation}

\subsection{Waterfall Region}
\label{sec:waterfall}
It was proved in \cite{AMRU04}, that the erasure probability due to large 
failures obeys a well defined scaling law. For our purpose it is best to
consider a refined scaling law which was conjectured 
in the same paper. For convenience of the reader we restate it here. 
\bconjecture[Refined Scaling Law]
\label{RefinedConjecture}
Consider transmission over a BEC of erasure probability $\epsilon$ using 
random elements from the ensemble 
$\eldpc n \ledge \redge= \nldpc n \lnodenormalized \rnodenormalized$.
Assume that the ensemble has a single critical point $y^*>0$ and 
let $\nu^*=\epsilon^* \lnodenormalized(y^*)$, 
where $\epsilon^*$ is the threshold erasure probability. 
Let $\prob^W_{\text{b}}(n, \ledge, \redge, \epsilon)$
(respectively, $\prob^W_{\text{B}}(n, \ledge, \redge, \epsilon)$)
denote the expected bit (block) erasure probability 
 {\em due to erasures of size
at least $n \gamma \nu^*$}, where $\gamma \in (0, 1)$.
Fix $z:=\sqrt{n} (\epsilon^*-\beta n^{-\frac{2}{3}}-\epsilon)$.
Then as $n$ tends to infinity,
\begin{align*}
\prob^W_{\text{B}}(n, \ledge, \redge, \epsilon)
& = Q\left(\frac{z}{\alpha} \right)\left(1+O(n^{-1/3}) \right), \\
\prob^W_{\text{b}}(n, \ledge, \redge, \epsilon)
& = \nu^* Q\left(\frac{z}{\alpha} \right)\left(1+O(n^{-1/3}) \right),
\end{align*}
where $\alpha=\alpha(\ledge, \redge)$ and $\beta=\beta(\ledge, \redge)$
are constants which depend on the ensemble.
\econjecture
In \cite{AMRU04,AMU05}, a procedure called covariance evolution was defined
to compute the scaling parameter $\alpha$ through the solution of a system 
of ordinary differential equations. 
The number of equations in the system is equal to the square of the number
of variable node degrees plus the largest check node degree minus one.
As an example, for an ensemble with $5$ different variable node degrees and 
$\drmax=30$, the number of coupled equations in covariance evolution is 
$(5+29)^2=1156$. The computation of the scaling parameter can therefore 
become a challenging task.
The main result in this paper is to show that it is possible to compute 
the scaling parameter $\alpha$ without explicitly solving covariance 
evolution. 
This is the crucial ingredient allowing for efficient code optimization.

\blemma[Expression for $\alpha$]
\label{expralphaLemma}
Consider transmission over a BEC with erasure probability $\epsilon$ using
random elements from the ensemble
$\eldpc n \ledge \redge= \nldpc n \lnodenormalized \rnodenormalized$.
Assume that the ensemble has a single critical point $y^*>0$, and let 
$\epsilon^*$ denote the  threshold erasure probability.
Then the scaling parameter $\alpha$ in Conjecture~\ref{RefinedConjecture} is given by
\begin{align*}
\alpha  = &
\left(
\frac{
\redge(\bar{x}^*)^2-\redge(\bar{x}^{*2})+
\redge'(\bar{x}^*)(1-2 x^* \redge(\bar{x}^*))-
\bar{x}^{*2} \redge'(\bar{x}^{*2})
}{
\lnodenormalized'(1) \ledge(y^*)^2 \redge'(\bar{x}^*)^2}\right.\\
&
+\left. \frac{
\epsilon^{*2} \ledge(y^*)^2-\epsilon^{*2} \ledge(y^{*2}) - y^{*2} \epsilon^{*2} \ledge'(y^{*2})}{
\lnodenormalized'(1) \ledge(y^*)^2} \right)^{1/2},
\end{align*}
where $x^*=\epsilon^* \ledge(y^*)$, $\bar{x}^*=1-x^*$. 
\elemma
The derivation of this expression is explained in 
Section~\ref{sec:alternative}

For completeness and the convenience of the reader, we repeat here also
an explicit characterization of the shift parameter $\beta$ which
appeared already (in a slightly different form) in \cite{AMRU04,AMU05}.
\bconjecture[Scaling Parameter $\beta$]
\label{exprbetaLemma}
Consider transmission over a BEC of erasure probability $\epsilon$ using
random elements from the ensemble
$\eldpc n \ledge \redge= \nldpc n \lnodenormalized \rnodenormalized$.
Assume that the ensemble has a single critical point $y^*>0$, and 
let $\epsilon^*$ denote the threshold erasure probability.
Then the scaling parameter $\beta$ in Conjecture~\ref{RefinedConjecture} is 
given by
\begin{eqnarray}
\label{equ:beta}
\beta/\Omega = &
\left(
\frac{\epsilon^{*4} r_2^{*2}(\epsilon^* \ledge'(y^*)^2 r_2^* - x^* (\ledge''(y^*) r_2^* + \ledge'(y^*) x^* ))^2}{\lnodenormalized'(1)^2 \redge'(\bar{x}^*)^3 x^{*10} (2 \epsilon^* \ledge'(y^*)^2 r_3^* - \ledge''(y^*) r_2^* x^*)}
\right)^{1/3},
\end{eqnarray}
where $x^*=\epsilon^* \ledge(y^*)$ and $\bar{x}^*=1-x^*$, and for $i \geq 2$
\begin{align*}
r_i^*  = &
\sum_{m \ge j \ge i} (-1)^{i+j}
\binom{j-1}{i-1} \binom{m-1}{j-1} \rho_m (\epsilon^* \lambda(y^*))^j\,.
\end{align*}
Further, $\Omega$ is a universal (code independent) constant 
defined in Ref.~\cite{AMRU04,AMU05}.
\econjecture
We also recall that  $\Omega$ is numerically quite close to $1$.
In the rest of this paper, we shall always adopt the approximate 
$\Omega$ by  $1$. 

\subsection{Error Floor}
\label{sec:errorfloor}
\blemma[Error Floor]
\label{exprerrorfloor}
Consider transmission over a BEC of erasure probability $\epsilon$ using 
random elements from an ensemble 
$\eldpc n \ledge \redge = \nldpc n \lnodenormalized \rnodenormalized$. Assume that
the ensemble has a single critical point $y^{*}>0$.
Let $\nu^* = \epsilon^* \lnodenormalized(y^*)$, where $\epsilon^*$
is the threshold erasure probability.
Let $\prob_{\text{b},\ssetsizemin}^E(n, \ledge, \redge, \epsilon)$
(respectively $\prob_{\text{B},\ssetsizemin}^E(n, \ledge, \redge, \epsilon)$)
denote the expected bit (block) erasure probability 
{\em due to stopping sets of size
between $\ssetsizemin$ and $n \gamma \nu^*$}, where $\gamma \in (0, 1)$.
Then, for any $\epsilon<\epsilon^*$,
\begin{align}
\prob_{\text{b},\ssetsizemin}^E(n, \ledge, \redge, \epsilon) = &\sum_{\ssetsize \geq \ssetsizemin} \ssetsize \tilde{A}_{\ssetsize}
\epsilon^{\ssetsize} \left(1+o(1)\right), \label{equ:approxb}\\
\prob_{\text{B},\ssetsizemin}^E(n, \ledge, \redge, \epsilon) = & 1 - e^{-\sum_{ \ssetsize\geq\ssetsizemin} \tilde{A}_{\ssetsize}\epsilon^{\ssetsize}}  
\left(1+o(1)\right), \label{equ:approxB}
\end{align}
where $\tilde{A}_{\ssetsize} = \coef{\log\left(A(x)\right)}{x^{\ssetsize}}$ for $\ssetsize \geq 1$, with $A(x)=\sum_{\ssetsize\geq 0} A_{\ssetsize} x^{\ssetsize}$ and
\begin{align}
A_s = &
\sum_{\edgesetsize} \left(
\coef{\prod_i (1+x y^i)^{n \lnodenormalized_i}}{x^{\ssetsize} y^{\edgesetsize}}\times \right. \label{equ:meanss}\\
& \left. \frac{\coef{\prod_i ((1+x)^i-ix)^{n (1-r) \rnodenormalized_i} }{x^e}}{\binom{n \lnodenormalized'(1)}
{\edgesetsize}}\right).\nonumber
\end{align}
\elemma
Discussion: In the lemma we only claim a multiplicative error term
of the form $o(1)$ since this is easy to prove. 
This weak statement would remain valid if we 
replaced the expression for $A_s$ given in (\ref{equ:meanss}) 
with the explicit and much easier
to compute asymptotic expression derived in \cite{RiU05}. 
In practice however the approximation is much better than the stated $o(1)$ 
error term
if we use the finite-length averages given by (\ref{equ:meanss}).
The hurdle in proving stronger error terms is due to the fact that
for a given length it is not clear how to relate 
the number of stopping sets to the number of {\em minimal} stopping sets. However, 
this relationship becomes easy in the limit of large blocklengths.

\begin{proof}
The key in deriving this erasure floor expression is in focusing
on the number of {\em minimal} stopping sets. These are stopping set
that are not the union of smaller stopping sets.
The {\em asymptotic} distribution of the
number of {\em minimal} stopping sets contained in an LDPC graph was already 
studied in \cite{RiU05}. We recall that the distribution
of the number of minimal stopping sets
tends to a Poisson distribution with 
independent components as the length tends to infinity. 
Because of this independence one can relate the number of minimal
stopping sets to the number of stopping sets -- any combination of
minimal stopping sets gives rise to a stopping set. In the limit of infinity
blocklenghts the minimal stopping sets are non-overlapping with probability
one so that the weight of the resulting stopping set is just the sum of
the weights of the individual stopping sets.
For example, the number of stopping sets of size two is equal
to the number of minimal stopping sets of size two plus the number of
stopping sets we get by taking all pairs of (minimal) stopping sets of size one. 

Therefore, define $\tilde{A}(x)=\sum_{\ssetsize\geq 1} \tilde{A}_{\ssetsize} x^{\ssetsize}$, 
with $\tilde{A}_{\ssetsize}$, the expected number of minimal stopping sets of
size $\ssetsize$ in the graph. Define further $A(x)=\sum_{\ssetsize\geq 0} A_{\ssetsize} x^{\ssetsize}$, with $A_{\ssetsize}$ the expected number of 
stopping sets of size $\ssetsize$ in the graph (not necessarily minimal).
We then have 
\begin{align*}
A(x)=e^{\tilde{A}(x)}=1+\tilde{A}(x)+\frac{\tilde{A}(x)^2}{2!}+\frac{\tilde{A}(x)^3}{3!}+\cdots,
\end{align*}
so that conversely $\tilde{A}(x)=\log\left( A(x)\right)$.

It remains to determine the number of stopping sets. 
As remarked right after the statement of the lemma, any expression which converges
in the limit of large blocklength to the asymptotic value would satisfy the statement of the lemma but we get
the best empirical agreement for short lengths 
if we use the exact finite-length averages.
These average were already compute in \cite{RiU05} and are given
as in (\ref{equ:meanss}).

Consider now e.g. the bit erasure probability. 
We first compute $A(x)$ using (\ref{equ:meanss}) and then
$\tilde{A}(x)$ by means of $\tilde{A}(x)=\log\left( A(x)\right)$.
Consider one minimal stopping set of size $s$. The probability that its
$s$ associated bits are all erased is equal to $\epsilon^s$ and if
this is the case this stopping set causes $s$ erasures. 
Since there are in expectation 
$\bar{A}_s$ minimal stopping sets of size $s$ and minimal stopping sets 
are non-overlapping with increasing probability 
as the blocklength increases
a simple union bound is asymptotically tight. The expression for the 
block erasure probability is derived in a similar way. Now we are interested
in the probability that a particular graph and noise realization results
in no (small) stopping set. Using the fact that the distribution of minimal
stopping sets follows a Poisson distribution we get equation ~(\ref{equ:approxB}).
\end{proof}

\subsection{Complete Approximation}
\label{sec:approximation}
In Section~\ref{sec:waterfall}, we have studied the erasure probability
stemming from failures of size bigger than $n\gamma\nu^*$ where $\gamma \in (0,1)$
and $\nu^*=\epsilon^* \lnodenormalized(y^*)$, i.e.,
$\nu^*$ is the asymptotic fractional number of erasures remaining after the 
decoding at the threshold. In Section~\ref{sec:errorfloor}, we have studied
the probability of erasures resulting from stopping sets of size between 
$\ssetsizemin$ and $n\gamma\nu^*$. 
Combining the results in the two previous sections, we get 
\begin{align}
\prob_B(n, \ledge, \redge, \epsilon) = & 
\prob_{B}^W(n, \ledge, \redge, \epsilon)+ \prob_{B,\ssetsizemin}^E(n, \ledge, \redge, \epsilon)\nonumber\\
= & Q\left(\frac{\sqrt{n} (\epsilon^*-\beta n^{-\frac{2}{3}}-\epsilon)}{\alpha} \right) \label{eq:OverallBlock}\\
&+ 1 - e^{-\sum_{ \ssetsize\geq\ssetsizemin} \tilde{A}_{\ssetsize}\epsilon^{\ssetsize}},\nonumber\\
&\nonumber\\
\prob_b(n, \ledge, \redge, \epsilon) = & \prob_{b}^W(n, \ledge, \redge, \epsilon)+ \prob_{b,\ssetsizemin}^E(n, \ledge, \redge, \epsilon)\nonumber\\
& \nu^* Q\left(\frac{\sqrt{n} (\epsilon^*-\beta n^{-\frac{2}{3}}-\epsilon)}{\alpha} \right) \label{eq:OverallBit}\\
& + \sum_{\ssetsize \geq \ssetsizemin} \ssetsize \tilde{A}_{\ssetsize}
\epsilon^{\ssetsize}. \nonumber
\end{align}

Here we assume that there is a single critical point.
If the degree distribution has several critical points
(at different values of the channel parameter $\epsilon^*_1$,
$\epsilon^*_2$,\dots) then we simply take
a sum of terms $\prob_B^W(n, \ledge, \redge, \epsilon)$, one for each
critical point.

Let us finally notice that summing the  probabilities of different 
error types provides in principle only an upper bound on the overall error
probability. However, for each given channel parameter $\epsilon$,
only one of the terms in Eqs.~(\ref{eq:OverallBlock}),
(\ref{eq:OverallBit}) dominates. As a consequence the bound is actually tight.

\section{Analytic Determination of $\alpha$}
\label{sec:alternative}
Let us now show how the scaling parameter $\alpha$ can be determined analytically.
We accomplish this in two steps. We first compute the variance of the number of erasure messages.
Then we show in a second step how to relate this variance to the scaling parameter $\alpha$.

\subsection{Variance of the Messages}
\begin{figure}[htp]
\centering
\setlength{\unitlength}{0.466bp}%
\begin{picture}(367.5, 231)
\put(0, 0){\includegraphics[scale=0.466]{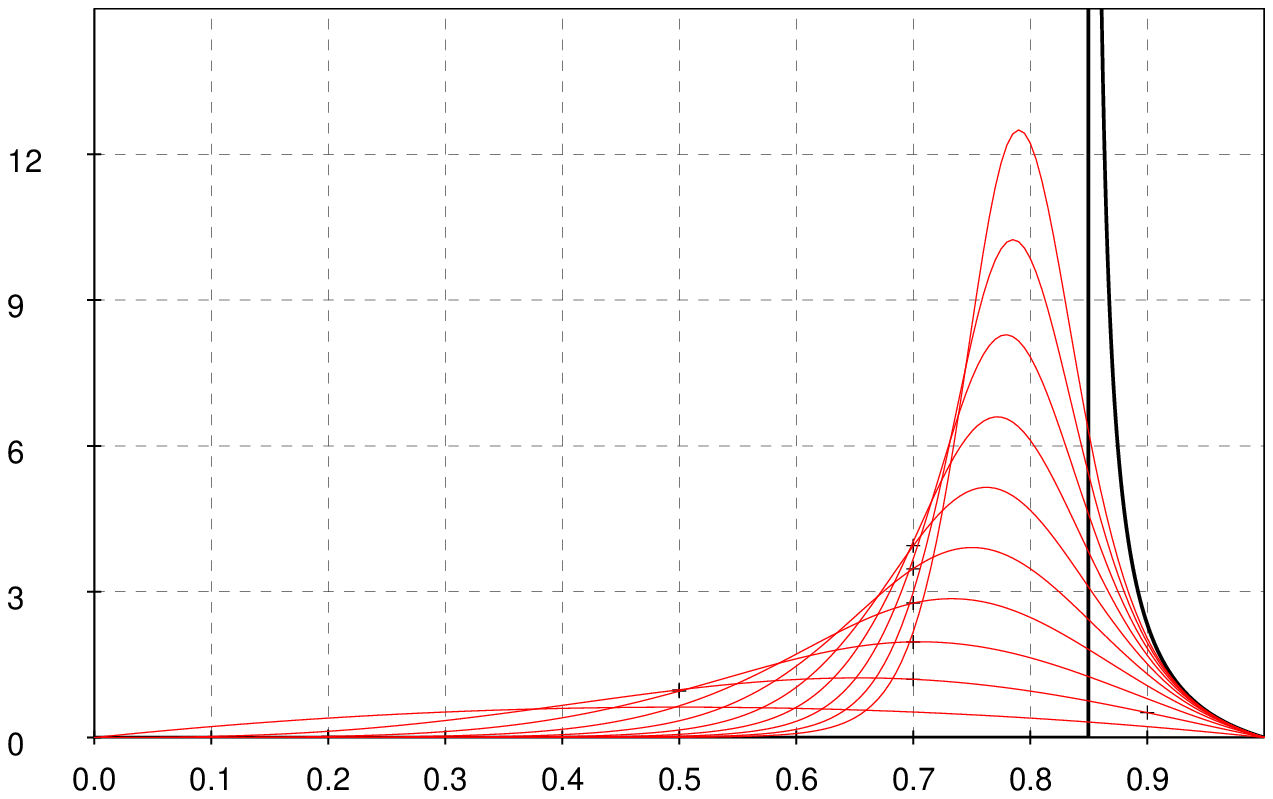}}
\end{picture}
\caption{
\label{fig:varianceforfinitel}
The variance as a function of $\epsilon$ and $\ell=0, \cdots, 9$ for
$(\lnodenormalized(x)=\frac25 x^2 + \frac35 x^3; \rnodenormalized(x)=\frac{3}{10} x^2+\frac{7}{10} x^3)$.}
\end{figure}

Consider the ensemble $\eldpc {n} {\ledge} {\redge}$ and assume that 
transmission takes place over a BEC of parameter $\epsilon$.
Perform $\ell$ iterations of BP decoding. Set $\mu_{i}^{(\ell)}$ equal to $1$ 
if the message sent out along edge $i$ from variable to check node is an 
erasure and $0$, otherwise.
Consider the variance of these messages in the limit of large blocklengths.
More precisely, consider
\begin{align*}
\variance^{(\ell)}\equiv
\lim_{n \rightarrow \infty} \frac{\expectation[(\sum_{i} \mu_{i}^{(\ell)})^2 ]-
\expectation[(\sum_{i} \mu_{i}^{(\ell)})]^2}{n \lnodenormalized'(1)}.
\end{align*}
Lemma \ref{lem:varianceforfinitel} in Section~\ref{sec:messagevariance} 
contains an analytic expression for this quantity as a function
of the degree distribution pair $(\ledge, \redge)$, the channel 
parameter $\epsilon$, and the number of iterations $\ell$. Let us consider
this variance as a function of the parameter $\epsilon$ and the number of
iterations $\ell$. Figure~\ref{fig:varianceforfinitel} shows the result
of this evaluation for the case 
$(\lnodenormalized(x)=\frac25 x^2 + \frac35 x^3; \rnodenormalized(x)=\frac{3}{10} x^2+\frac{7}{10} x^3)$.
The threshold for this example is 
$\epsilon^* \approx  0.8495897455$. This value is indicated as 
a vertical line in the figure. As we can see from this figure, the
variance is a unimodal function of the channel parameter.
It is zero for the extremal values of $\epsilon$ (either
all messages are known or all are erased) and it takes
on a maximum value for a parameter of $\epsilon$ which approaches
the critical value $\epsilon^*$ as $\ell$ increases.
Further, for increasing $\ell$ the maximum value of the variance increases.
The limit of these curves
as $\ell$ tends to infinity 
$\variance = \lim_{\ell\to\infty}\variance^{(\ell)}$
is also shown (bold curve): 
the variance is zero below  threshold; above  threshold
it is positive and diverges as the threshold is approached.
In Section \ref{sec:messagevariance} we state
the exact form of the limiting curve. We show that for $\epsilon$ 
approaching $\epsilon^*$ from above
\begin{align}
\variance = \frac{\gamma}{(1- \epsilon \ledge'(y) \redge'(\bar{x}))^2} + 
O((1- \epsilon \ledge'(y) \redge'(\bar{x}))^{-1}),
\label{eq:VarianceDivergence} 
\end{align}
where
\begin{align*}
\gamma &=  \epsilon^*{}^2 \ledge'(y^*)^2\left\{[
\redge(\bar{x}^*)^2-\redge(\bar{x}^*{}^2)+
\redge'(\bar{x}^*)(1-2 x^* \redge(\bar{x}^*))
\right.
\\
-&\left.\bar{x}^*{}^2 \redge'(\bar{x}^*{}^2)]+\epsilon^*{}^2
\redge'(\bar{x}^*)^2 [
 \ledge(y^*)^2-\ledge(y^*{}^2) - y^*{}^2 \ledge'(y^*{}^2)]\right\}\, .
\end{align*}
Here $y^*$ is the unique critical point, $x^*=\epsilon^* \ledge(y^*)$,
and $\bar{x}^*=1-x^*$. Since $(1- \epsilon \ledge'(y) \redge'(\bar{x}))=
\Theta(\sqrt{\epsilon-\epsilon^*})$, Eq.~(\ref{eq:VarianceDivergence})
implies a divergence at $\epsilon^*$.

\subsection{Relation Between $\gamma$ and $\alpha$}
\begin{figure}[htp]
\centering
\setlength{\unitlength}{0.466bp}%
\begin{picture}(367.5, 231)
\put(0, 0){\includegraphics[scale=0.466]{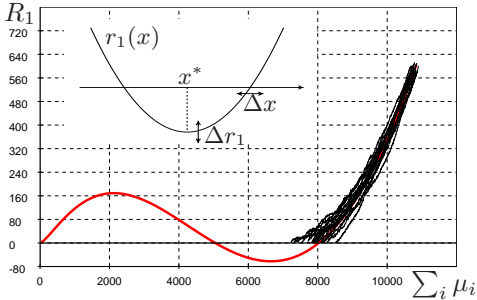}}
\put(357,5){\makebox(0,0){$\sum_{i} \mu_{i}$}}
\put(14,225){\makebox(0,0){$R_1$}}

\put(180,126){\makebox(0,0){\small $\Delta r_1$}}
\put(208,152.5){\makebox(0,0){\small $\Delta x$}}
\put(105,205){\makebox(0,0){\small $r_1(x)$}}
\put(152,175){\makebox(0,0){\small $x^*$}}

\end{picture}
\caption{\label{fig:curvesabove}
Number of degree-one check nodes as a function of the number of erasure messages
in the corresponding BP decoder for 
$\nldpc{n=8192}{\ledge(x)=x^2}{\redge(x)=x^5}$. The thin lines represent the
decoding trajectories that stop when $r_1=0$ and the thick line is the mean 
curve predicted by density evolution.} 
\end{figure}

Now that we know  the asymptotic variance of the edges messages, let us discuss
how this quantity can be related to the scaling parameter $\alpha$.
Think of a decoder operating above the threshold of the code. Then, for large 
blocklengths, it will get stuck with high probability before correcting all nodes.
In Fig~\ref{fig:curvesabove} we show $R_1$, the number of degree-one 
check nodes, as a function of the number of erasure messages
for a few decoding runs.
Let $\variance_*$ represent the normalized variance of the number of
erased messages in the decoder after an infinite number of iterations
\begin{align*}
\variance_*\equiv
\lim_{n \rightarrow \infty} \lim_{\ell\to\infty}\frac{\expectation[(\sum_{i} \mu_{i}^{(\ell)})^2 ]-
\expectation[(\sum_{i} \mu_{i}^{(\ell)})]^2}{n \lnodenormalized'(1)}.
\end{align*}
In other words, $\variance_*$ is
the variance of the point at which the decoding trajectories 
hit the $R_1=0$ axis.

This quantity can be related to the variance of the number of degree-one
check nodes through the slope of the density evolution curve. 
Normalize all the quantities by $n\lnodenormalized'(1)$,
 the number of edges in the graph.
Consider the  curve $r_1(\epsilon, x)$ given by density evolution, and
representing the fraction of degree-one check nodes in the residual graph, 
around the critical point for an erasure probability above the threshold 
(see Fig.\ref{fig:curvesabove}). The real decoding process stops when 
hitting the $r_1=0$ axis. Think of a virtual process identical to the 
decoding for $r_1>0$ but that continues below the $r_1=0$ axis 
(for a proper definition, see \cite{AMRU04}).
A simple calculation shows that if the point at which the curve hits the 
x-axis varies by $\Delta x$ while keeping the minimum at $x^*$, 
it results in a variation of the height of the curve by
\begin{align*}
\Delta r_1  = & \frac{\partial^2 r_1(\epsilon, x)}{\partial x^2}\Big|_{*} (x-x^{*}) \Delta x + o(x- x^{*})
\end{align*}
Taking the expectation of the square on both side and letting $\epsilon$ tend
to $\epsilon^*$, we obtain the normalized variance of $R_1$ at threshold
\begin{align*}
\delta^{r_1, r_1}|_*
 = & \lim_{\epsilon \rightarrow \epsilon^*} \left( \left( \frac{\partial^2 r_1(\epsilon, x)}{\partial x^2}\Big|_* \right)^2 (x-x^*)^2 \variance + o((x-x^*)^2)
\right)\\
 = & \left(\frac{x^*}{\epsilon^* \ledge'(y^*)} \right)^2 
\lim_{\epsilon \rightarrow \epsilon^*} 
(1-\epsilon \ledge'(y) \redge'(\bar{x}))^2 \variance_*\, .
\end{align*}
The transition between the first and the 
second line comes the relationship between the $\epsilon$ and $x$, with 
$r_1(\epsilon,x)=0$ when $\epsilon$ tends to $\epsilon^*$.

The quantity $\variance_*$ differs from $\variance$ computed in the previous 
paragraphs because of the different order of the limits
$n\to\infty$ and $\ell\to\infty$. However it can be proved that
the order does not matter and $\variance=\variance_*$.
Using the result (\ref{eq:VarianceDivergence}), we finally get
\begin{align*}
\delta^{r_1, r_1}|_*
= \left(\frac{x^*}{\epsilon^* \ledge'(y^*)} \right)^2 \gamma\, .
\end{align*}
We conclude that the scaling parameter $\alpha$ can be obtained as 
\begin{align*}
\alpha = & \sqrt{\frac{\delta^{r_1 r_1}|_*}{\lnodenormalized'(1)
\left( \frac{\partial r_1}{\partial \epsilon} \right)^2}} 
= \sqrt{\frac{\gamma}{\lnodenormalized'(1) x^{*2} \ledge'(y^*)^2 \redge'(\bar{x}^*)^2}} 
\end{align*}
The last expression is equal to the one in Lemma \ref{expralphaLemma}.

\section{Message Variance}
\label{sec:messagevariance}
Consider the ensemble $\nldpc n \ledge \redge$ and transmission over the BEC 
of erasure probability $\epsilon$. As pointed out in the previous section,
the scaling parameter $\alpha$ can be related to the (normalized) 
variance, with respect to the choice of the graph and the channel realization, 
of the number of erased edge messages sent from the variable nodes.
Although what really matters is the limit of this quantity as the blocklength 
and the number of iterations tend to infinity (in this order) 
we start by providing an exact expression for finite number of iterations 
$\ell$ (at infinite blocklength). At the end of this section, we shall
take the limit $\ell\to\infty$.

To be definite, we initialize the iterative decoder by setting all 
check-to-variable messages to be erased at time $0$. 
We let $x_i$ (respectively $y_i$) be the fraction of 
erased messages sent from variable to  check nodes 
(from check to  variable nodes),
at iteration $i$, in the infinite blocklength limit.
These values are determined by the density evolution \cite{RiU05} 
recursions $y_{i+1}=1-\redge(\bar{x}_{i})$, with $x_i=\epsilon 
\ledge(y_i)$ (where we used the notation $\bar{x}_i = 1-x_i$). 
The above initialization  implies $y_0=1$. For future convenience 
we also set $x_i=y_i=1$ for $i<0$.

Using these variables, we have the following characterization of 
$\variance^{(\ell)}$, the (normalized) variance after $\ell$ 
iterations. 
\blemma
\label{lem:varianceforfinitel}
Let $\graph$  be chosen uniformly at random from $\nldpc n \ledge \redge$ and 
consider transmission over the BEC of erasure probability $\epsilon$. Label the 
$n \lnodenormalized'(1)$ edges of $\graph$ in some fixed order by the elements 
of $\{1,\cdots,n \lnodenormalized'(1)\}$. Assume that the receiver performs 
$\ell$ rounds of Belief Propagation decoding and let $\mu^{(\ell)}_i$ be equal 
to one if the message sent at the end of the $\ell$-th iteration along edge $i$
(from a variable node to a check node) is an erasure, and zero otherwise. Then
\begin{align}
\variance^{(\ell)} \equiv & \lim_{n \rightarrow \infty}
\frac{\expectation[(\sum_{i} \mu_{i}^{(\ell)})^2 ]-
\expectation[(\sum_{i} \mu_{i}^{(\ell)})]^2}{n \lnodenormalized'(1)} \label{equ:finitel}\\
= &
x_{\ell}+
x_{\ell} (1, 0) \biggl(\sum_{j=0}^{\ell} 
\Vtree(\ell) 
\cdots 
\Ctree(\ell-j)
\biggr) (1, 0)^T  \tag*{edges in $\tree_1$} \\
& + x_{\ell}^2 \redge'(1) \sum_{i=0}^{\ell-1} \ledge'(1)^i \redge'(1)^{i} \tag*{edges in $\tree_2$}\\
& +
x_{\ell} (1, 0) \biggl(\sum_{j=1}^{2 \ell} 
\Vtree(\ell) 
\cdots 
\Ctree(\ell-j)\biggr) (1, 0)^T  \tag*{edges in $\tree_3$}\\
& + 
(1,0)
\biggl(\sum_{j=0}^{\ell}
\left(y_{\ell-j}U^{\star}(j, j) + 
(1-y_{\ell-j}) U^{0}(j, j) \right)\biggr. \nonumber \\ 
& \quad \quad +\sum_{j=\ell+1}^{2 \ell}
\Vtree(\ell) 
\cdots 
\Ctree(2 \ell-j) \nonumber \\ 
& \biggl. \phantom{\sum_{j=0}^{\ell}}  \quad \quad \quad 
\cdot \left(y_{\ell-j}U^{\star}(j, \ell) + 
(1-y_{\ell-j}) U^{0}(j, \ell) \right) \biggr)-
\tag*{edges in $\tree_4$} 
\end{align}
\begin{align}
\phantom{\variance^{(\ell)} \equiv } & - x_{\ell} W(\ell,1)
\nonumber
\\
& + \sum_{i=1}^{\ell} F_i 
\left(x_i W(\ell,1)
-\epsilon W(\ell,y_i)
\right)\nonumber\\
& - \sum_{i=1}^{\ell} F_i \epsilon \ledge'(y_i) \bigl(
D(\ell,1)
\redge(\bar{x}_{i-1}) 
- D(\ell,\bar{x}_{i-1}
\bigr) \nonumber\\
& + \sum_{i=1}^{\ell-1} F_i \left(x_{\ell}+
(1, 0)
\Vtree(\ell) 
\cdots 
\Ctree(0)\Vtree(0)(1,0)^T\right)\nonumber\\
& \quad \quad  \quad  \quad \cdot 
(1, 0) \Vtree(i) 
\cdots 
\Ctree(i-\ell) (1, 0)^T\nonumber\\
& - \sum_{i=1}^{\ell-1} F_i  x_i \left(x_{\ell}+
(1, 0)\Vtree(\ell) 
\cdots 
\Ctree(0)\Vtree(0)(1,0)^T\right), \nonumber\\
& \quad \quad \quad \quad \quad \quad \cdot \left(\ledge'(1)\redge'(1)\right)^{\ell} \nonumber
\end{align}
where we introduced the shorthand
\begin{eqnarray}
\Vtree(i)\cdots\Ctree(i-j) \equiv \prod_{k=0}^{j-1}\Vtree(i-k)\Ctree(i-k-1).
\end{eqnarray} We
%
We define the matrices
\begin{align}
\Vtree(i) & = 
\left(
\begin{array}{cc}
\epsilon \ledge'(y_i) & 0 \\
\ledge'(1)-\epsilon \ledge'(y_i) & \ledge'(1)
\end{array}
\right), \label{equ:Vtree}\\ 
\Ctree(i) & =
\left(
\begin{array}{cc}
\redge'(1) & \redge'(1)-\redge'(\bar{x}_i) \\
0 & \redge'(\bar{x}_i)
\end{array}
\right), & i \geq 0,\label{equ:Ctree}\\
&\nonumber\\
\Vtree(i) & = 
\left(
\begin{array}{cc}
\ledge'(1) & 0 \\
0 & \ledge'(1)
\end{array}
\right),\label{equ:Vtreeneg} \\ 
\Ctree(i) & =
\left(
\begin{array}{cc}
\redge'(1) & 0 \\
0 & \redge'(1)
\end{array}
\right), & i < 0 \label{equ:Ctreeneg}.
\end{align}
Further,
$U^{\star}(j,j)$, $U^{\star}(j,\ell)$, $U^{0}(j,j)$ and $U^{0}(j,\ell)$ are computed through the following recursion.
For $j\leq \ell$, set
\begin{align*}
U^{\star}(j,0) = & (y_{\ell-j} \epsilon \ledge'(y_{\ell}), (1-y_{\ell-j})  \epsilon \ledge'(y_{\ell}))^T,\\
U^{0}(j,0) = & (0,0)^T,
\end{align*}
whereas for $j>\ell$, initialize by
\begin{align*}
U^{\star}(j,j-\ell) = & 
(1, 0) 
\Vtree(\ell) 
\cdots 
\Ctree(2\ell-j) (1, 0)^T \binom{\epsilon\ledge'(y_{2\ell-j})}{0}\\
& \!\!\!\!\!\!\!\!\!\!\!\!\!\!\!\!\!\!\!\!\!\!\!\!\!\!\!\!+ (1, 0) \Vtree(\ell) 
\cdots 
\Ctree(2\ell-j) (0, 1)^T \binom{\epsilon(\lambda'(1)-\lambda'(y_{2\ell-j}))}{
\lambda'(1)(1-\epsilon)} ,\\
U^{0}(j,j-\ell) = &  (1, 0) \Vtree(\ell) 
\cdots 
\Ctree(2\ell-j) (0, 1)^T
\binom{\epsilon\lambda'(1)}{(1-\epsilon)\lambda'(1)} .
\end{align*}
The recursion is 
\begin{align}
U^{\star}(j, k) = & M_1(j,k) \Ctree(\ell-j+k-1) U^{\star}(j, k-1)\label{eq:RecT4_1}\\
& + M_2(j,k) [  N_1(j,k-1) U^{\star}(j, k-1)\nonumber \\
&\quad \quad \quad \quad \quad \quad \quad \quad \quad  
+ N_2(j,k-1) U^{0}(j, k-1) ] , \nonumber\\
U^{0}(j, k) = & \Vtree(\ell-j+k) [  N_1(j,k-1) U^{\star}(j, k-1) 
\label{eq:RecT4_2}\\
&\quad \quad \quad \quad \quad \quad \quad \quad \quad
+ N_2(j,k-1) U^{0}(j, k-1) ],\nonumber
\end{align}
with
\begin{align*}
& M_1(j,k) =  \\
& \quad \left(
\begin{array}{cc}
\epsilon \ledge'( y_{\max\{\ell - k, \ell-j+k\}} ) & 0 \\
\ind_{\{j<2 k\}} \epsilon (\ledge'(y_{\ell-k})-\ledge'(y_{\ell-j+k})) & \epsilon \ledge'(y_{\ell-k})
\end{array}
\right), \\
& M_2(j,k) = \\
& \quad \left(
\begin{array}{cc}
\ind_{\{j>2 k\}} \epsilon (\ledge'(y_{\ell-j+k})-\ledge'(y_{\ell-k})) & 0 \\
\ledge'(1)-\epsilon \ledge'(y_{\min\{\ell-k, \ell-j+k \}}) & \ledge'(1)-\epsilon \ledge'(y_{\ell-k})
\end{array}
\right), \\
& N_1(j,k) = \\ 
& \quad \left(
\begin{array}{cc}
\redge'(1)-\redge'(\bar{x}_{\ell-k-1}) & \redge'(1)-
\redge'(\bar{x}_{\max\{\ell-k-1, \ell-j+k\}}) \\
0 & \ind_{\{j\leq 2 k\}}(\redge'(\bar{x}_{\ell-j+k})-
\redge'(\bar{x}_{\ell-k-1}))
\end{array}
\right), \\
& N_2(j,k) = \\ 
& \quad \left(
\begin{array}{cc}
\redge'(\bar{x}_{\ell-k-1}) & \ind_{\{j>2 k\}}(\redge'(\bar{x}_{\ell-k-1})-\redge'(\bar{x}_{\ell-j+k})) \\
0 & \redge'(\bar{x}_{\min\{\ell-k-1, \ell-j+k\}}) 
\end{array}
\right).
\end{align*}

The coefficients $F_i$ are given by
\begin{align}
F_i = & \prod_{k=i+1}^{\ell}\epsilon \ledge'(y_{k}) \redge'(\bar{x}_{k-1}),\label{equ:Fi}
\end{align}
and finally
\begin{align*}
W(\ell,\alpha) = & 
\sum_{k=0}^{2 \ell} 
(1, 0)\Vtree(\ell) 
\cdots 
\Ctree(\ell-k) 
A(\ell, k, \alpha) \\
& + x_{\ell} (\alpha \ledge'(\alpha)+\ledge(\alpha)) \rho'(1)
\sum_{i=0}^{\ell-1} \redge'(1)^i \ledge'(1)^{i},
\end{align*}
with $A(\ell, k, \alpha)$ equal to
\begin{align*}
&\left(
\begin{array}{c}
\epsilon \alpha y_{\ell-k} \ledge'(\alpha y_{\ell-k})+
\epsilon \ledge(\alpha y_{\ell-k}) \\
\alpha \ledge'(\alpha)+ \ledge(\alpha)-
\epsilon \alpha y_{\ell-k} \ledge'(\alpha y_{\ell-k})-
\epsilon \ledge(\alpha y_{\ell-k}) \\
\end{array}
\right), 
\tag*{$k \leq \ell$} \\
&\left(
\begin{array}{c}
\alpha \ledge'(\alpha)+
\ledge(\alpha) \\
0
\end{array}
\right), \tag*{$k > \ell$}
\end{align*}
and
\begin{align*}
& D(\ell,\alpha)  = 
\sum_{k=1}^{2 \ell} 
(1, 0) \Vtree(\ell) 
\cdots 
\Ctree(\ell-k+1)\Vtree(\ell-k+1) \\
& \quad \cdot \left(
\begin{array}{c}
\alpha \redge'(\alpha)+\redge(\alpha)-
\alpha(\bar{x}_{\ell-k}) \redge'(\alpha \bar{x}_{\ell-k})-
\redge(\alpha \bar{x}_{\ell-k}) \\
\alpha \bar{x}_{\ell-k} \redge'(\alpha \bar{x}_{\ell-k})+
\redge(\alpha \bar{x}_{\ell-k})
\end{array}
\right) \\
& + x_{\ell} (\alpha \redge'(\alpha)+\redge(\alpha))
\sum_{i=0}^{\ell-1} \redge'(1)^{i} \ledge'(1)^{i}.
\end{align*}
\elemma

\begin{proof}
Expand $\variance^{(\ell)}$ in (\ref{equ:finitel}) as
\begin{align}
\variance^{(\ell)} = & \lim_{n \rightarrow \infty}
\frac{\expectation[(\sum_{i} \mu_{i}^{(\ell)})^2 ]-
\expectation[\sum_{i} \mu_{i}^{(\ell)}]^2}{n \lnodenormalized'(1)}, \nonumber\\
= & \lim_{n \rightarrow \infty}
\frac{\sum_j \left( \expectation[\mu_{j}^{(\ell)}\sum_{i} \mu_{i}^{(\ell)}]-
\expectation[\mu_{j}^{(\ell)}] \expectation[\sum_{i} \mu_{i}^{(\ell)}]\right)}{n \lnodenormalized'(1)}, \nonumber\\
= & \lim_{n \rightarrow \infty}
\expectation[\mu_{1}^{(\ell)}\sum_{i} \mu_{i}^{(\ell)}]-
\expectation[\mu_{1}^{(\ell)}] \expectation[\sum_{i} \mu_{i}^{(\ell)}], 
\nonumber\\
= & \lim_{n \rightarrow \infty}
\expectation[\mu_{1}^{(\ell)}\sum_{i} \mu_{i}^{(\ell)}]- n \lnodenormalized'(1) x_{\ell}^2.\label{equ:mu1}
\end{align}
In the last step, we have used the fact that $x_{\ell}=\expectation[\mu_{i}^{(\ell)}]$ for any $i\in\{1,\cdots,\lnode'(1)\}$.
Let us look more carefully at the first term of (\ref{equ:mu1}).
After a finite number of iterations, each message $\mu_i^{(\ell)}$ 
depends upon the received symbols of a {\em subset} of the variable nodes.
Since $\ell$ is kept finite, this subset remains finite in the
large blocklength limit, and by standard arguments is a tree with 
high probability. As usual,
we refer to the subgraph containing all such variable nodes,
as well as the check nodes connecting them as to the {\em computation  tree}
for $\mu^{(\ell)}_i$.

It is useful to split the sum in the first term of Eq.~(\ref{equ:mu1})
into two contributions: the first contribution stems from edges $i$ so that the computation
trees of $\mu^{(\ell)}_1$ and  $\mu^{(\ell)}_i$ {\em intersect}, and the second one
stems from the remaining edges.
More precisely, we write
\begin{align}
&\lim_{n \rightarrow \infty} \left(\expectation[\mu_{1}^{(\ell)}\sum_{i} \mu_{i}^{(\ell)}] - n \lnodenormalized'(1) x_{\ell}^2\right) =  \lim_{n \rightarrow \infty} \expectation[\mu_{1}^{(\ell)}\sum_{i\in\tree} \mu_{i}^{(\ell)}]\nonumber \\
& + \lim_{n \rightarrow \infty} \left(\expectation[\mu_{1}^{(\ell)}\sum_{i\in\btree} \mu_{i}^{(\ell)}]  - n \lnodenormalized'(1) x_{\ell}^2  \right).\label{equ:treeornot}
\end{align}
We define $\tree$ to be that subset of the variable-to-check edge indices so
that if $i \in \tree$ then
the computation trees $\mu_{i}^{(\ell)}$  and $\mu_1^{(\ell)}$ intersect. 
This means that $\tree$ includes all the edges
whose messages  depend on some of the received values that
are used in the computation of $\mu_1^{(\ell)}$. For convenience, we complete $\tree$
by including all edges that are connected to the same variable nodes as 
edges that are already in $\tree$.
$\btree$ is the complement in $\{1,\cdots,n \lnodenormalized'(1)\}$ of the set 
of indices $\tree$.

\begin{figure}[htp]
\centering
\setlength{\unitlength}{0.3bp}%
\begin{picture}(600, 640)
\put(0, 0){\includegraphics[scale=0.3]{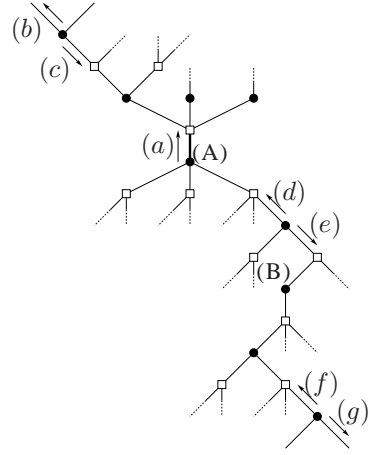}}
\put(260,420){\makebox(0,0){$(a)$}}
\put(95,565){\makebox(0,0){$(b)$}}
\put(130,515){\makebox(0,0){$(c)$}}
\put(425,360){\makebox(0,0){$(d)$}}
\put(470,320){\makebox(0,0){$(e)$}}
\put(465,120){\makebox(0,0){$(f)$}}
\put(505,85){\makebox(0,0){$(g)$}}
\put(325,410){\makebox(0,0){\small \sc(A)}}
\put(405,260){\makebox(0,0){\small \sc(B)}}
\end{picture}
\caption{\label{fig:tree}
Graph representing all edges contained in $\tree$, for the case of $\ell=2$. 
The small letters represent messages sent along the edges from a variable node 
to a check node and the capital letters represent variable nodes.
The message $\mu_1^{(\ell)}$ is represented by $(a)$.}
\end{figure}

The set of indices $\tree$ depends on the number of iterations performed and
on the graph realization. For any fixed $\ell$, $\tree$ is a tree with 
high probability in the large blocklength limit, and admits a simple 
characterization. It contains two sets of edges: the ones `above' and 
the ones `below' edge $1$ (we call this the `root' edge and the variable 
node it is connected to, the `root' variable node). 
Edges \emph{above} the root 
are the ones departing from a variable node that can be reached 
by a non reversing path starting with the root edge and involving at most 
$\ell$ variable nodes (not including the root one). Edges \emph{below}
the root are the ones departing from a variable node that can be reached 
by a non reversing path starting with the opposite of the root edge and 
involving at most $2\ell$ variable nodes (not including the root one). 
Edges departing from the root variable node are considered below the
root (apart from the root edge itself).

We have depicted in Fig.~\ref{fig:tree}
an example for the case of an irregular graph with $\ell=2$.
In the middle of the figure the edge $(a)\equiv 1$ carries the message 
$\mu_1^{(\ell)}$. We will call  
$\mu_1^{(\ell)}$ the root message. We expand the graph starting from this 
root node. We consider $\ell$ variable node levels above the root. 
As an example, notice that the channel output on node {\sc (A)} affects 
$\mu_1^{(\ell)}$ as well as the message sent on $(b)$ at the 
$\ell$-th iteration.
Therefore the corresponding computation trees intersect and,
according to our definition $(b)\in \tree$. On the other hand, the 
computation tree of $(c)$ does not intersect the one of $(a)$, 
but $(c)\in \tree$ because it shares a variable node with $(b)$.
We also expand $2\ell$ levels below the root. 
For instance, the value received on node {\sc (B)} affects both 
$\mu_1^{(\ell)}$ and the message sent on $(g)$ at the  $\ell$-th iteration.

We compute the two terms in (\ref{equ:treeornot}) separately. Define
$S= \lim_{n \rightarrow \infty} \expectation[\mu_{1}^{(\ell)}\sum_{i\in\tree} 
\mu_{i}^{(\ell)}]$ and 
$S^c=\lim_{n \rightarrow \infty} \left(\expectation[\mu_{1}^{(\ell)}\sum_{i\in\btree} \mu_{i}^{(\ell)}]  - n \lnodenormalized'(1) x_{\ell}^2  \right)$.
\begin{figure}[htp]
\centering
\setlength{\unitlength}{0.2bp}%
\begin{picture}(600, 1000)
\put(0, 0){\includegraphics[scale=0.2]{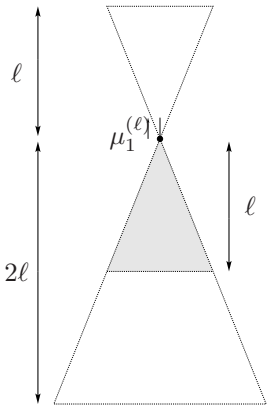}}
\put(30,350){\makebox(0,0){$2\ell$}}
\put(30,725){\makebox(0,0){$\ell$}}
\put(245,610){\makebox(0,0){$\mu_1^{(\ell)}$}}
\put(470,475){\makebox(0,0){$\ell$}}
\end{picture}
\caption{\label{fig:treesize}
Size of $\tree$. It contains $\ell$ layers of variable nodes above the root edge
and $2\ell$ layer of variable nodes below the root variable node. The gray 
area represent the computation tree of the message $\mu_1^{(\ell)}$. It 
contains $\ell$ layers of variable nodes below the root variable node.}
\end{figure}

\subsubsection{Computation of $S$}
Having defined $\tree$, we can further identify four different types of terms
appearing in $S$ and write
\begin{align*}
S = & \lim_{n \rightarrow \infty} \expectation[\mu_{1}^{(\ell)}\sum_{i\in\tree} \mu_{i}^{(\ell)}]\\
= & \lim_{n \rightarrow \infty} \expectation[\mu_{1}^{(\ell)}\sum_{i\in\tree_1} \mu_{i}^{(\ell)}] + 
\lim_{n \rightarrow \infty} \expectation[\mu_{1}^{(\ell)}\sum_{i\in\tree_2} \mu_{i}^{(\ell)}] + \\
& \lim_{n \rightarrow \infty} \expectation[\mu_{1}^{(\ell)}\sum_{i\in\tree_3} \mu_{i}^{(\ell)}] + \lim_{n \rightarrow \infty} \expectation[\mu_{1}^{(\ell)}\sum_{i\in\tree_4} \mu_{i}^{(\ell)}]
\end{align*}
The subset $\tree_1 \subset \tree$ contains the edges above the root 
variable node that carry messages that point upwards 
(we include the root edge in $\tree_1$).
In Fig.~\ref{fig:tree}, the message sent on edge $(b)$ is of this type. 
$\tree_2$ contains all edges above the root but point downwards,
such as $(c)$ in Fig.~\ref{fig:tree}.
$\tree_3$ contains the edges  below the root that carry an upward
messages, like $(d)$ and $(f)$. Finally, $\tree_4$ 
contains the edges below the root variable node that point downwards, 
like $(e)$ and $(g)$.

Let us start with the simplest term, involving the messages in $\tree_2$.
If $i\in\tree_2$, then the computation trees of 
$\mu_1^{(\ell)}$, and $\mu_i^{(\ell)}$ are with high probability 
disjoint in the large blocklength limit. In this case, the messages
$\mu_1^{(\ell)}$ and $\mu_i^{(\ell)}$ do not depend on any
common channel observation. The messages are nevertheless correlated: 
conditioned on the computation graph of the root edge the degree distribution of
the computation graph of edge $i$ is biased (assume that the computation graph
of the root edge
contains an unusual number of high degree check nodes; then the computation graph of
edge $i$ must contain in expectation an unusual low number of high degree check nodes).
This correlation is however of order $O(1/n)$ and since $\tree$ only contains a finite
number of edges the contribution of this correlation vanishes as $n\to\infty$.
We obtain therefore
\begin{align*}
\lim_{n \rightarrow \infty} \expectation[\mu_{1}^{(\ell)}\sum_{i\in\tree_2} \mu_{i}^{(\ell)}] = &
x_{\ell}^2 \redge'(1) \sum_{i=0}^{\ell-1} \redge'(1)^i \ledge'(1)^{i},
\end{align*}
where we used $ \lim_{n \rightarrow \infty} \expectation[\mu_{1}^{(\ell)}\mu_{i}^{(\ell)}] = x_{\ell}^2$, and the fact that 
the expected number of edges in $\tree_2$ is 
$\redge'(1) \sum_{i=0}^{\ell-1} \ledge'(1)^i \redge'(1)^{i}$.

For the edges in $\tree_1$ we obtain
\begin{align}
& \lim_{n \rightarrow \infty} \expectation[\mu_{1}^{(\ell)}\sum_{i\in\tree_1} \mu_{i}^{(\ell)}] = 
x_{\ell}+ \label{equ:tree1}\\
& x_{\ell} (1, 0)\left(\sum_{j=1}^{\ell} \Vtree(\ell) \Ctree(\ell-1) \cdots \Vtree(\ell-j+1) \Ctree(\ell-j)\right) (1, 0)^T, \nonumber
\end{align}
with the matrices $\Vtree(i)$ and $\Ctree(i)$ defined in 
Eqs.~(\ref{equ:Vtree}) and (\ref{equ:Ctree}).
In order to understand this expression, consider the following case
(cf. Fig.~\ref{fig:CVtree} for an illustration). We are 
at the $i$-{th} iteration of BP decoding and we pick an edge at random in the 
graph. It is connected to a check node of degree $j$ with probability
$\redge_j$. Assume further that the message carried by 
this edge from the variable node to the check node (incoming message) is erased 
with probability $p$ and known with probability $\bar{p}$. We want to compute 
the expected numbers of erased and known messages sent out by the check
node on its other edges (outgoing messages).
If the incoming message is erased, then the number of erased outgoing messages
is exactly $(j-1)$. Averaging over the check node degrees 
gives us $\redge'(1)$. If the incoming message is known, then the expected 
number of erased outgoing messages is $(j-1) (1-(1-x_i)^{j-2})$. Averaging 
over the check node degrees gives us $\redge'(1)-\redge'(1-x_i)$.
The expected number of erased outgoing messages is therefore, 
$p \redge'(1) + \bar{p} (\redge'(1)-\redge'(1-x_i))$. 
Analogously,
the expected number of known outgoing messages is $\bar{p} \redge'(x_i)$.
This result can be written using a matrix notation: 
the expected number of erased (respectively, known) outgoing messages is the 
first (respectively, second) component of the vector
$\Ctree(i)(p,\bar{p})^T$, with $\Ctree(i)$ being defined in 
Eq.~(\ref{equ:Ctree}).

The situation is similar if we consider a variable node instead of the check
node with the matrix the matrix $\Vtree(i)$ replacing $\Ctree(i)$. 
The result is generalized to several layers of check and variable
nodes, by taking the product of the corresponding matrices, 
cf. Fig.~\ref{fig:CVtree}.

\begin{figure}[htp]
\centering
\setlength{\unitlength}{0.45bp}%
\begin{picture}(480, 240)
\put(0, 0){\includegraphics[scale=0.45]{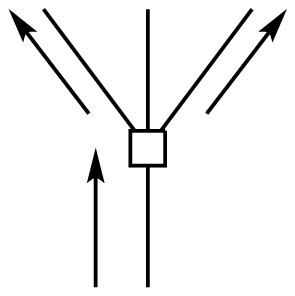}}
\put(160, 0){\includegraphics[scale=0.45]{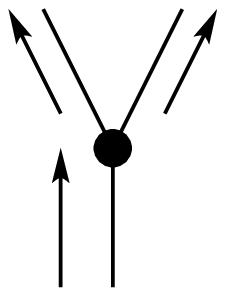}}
\put(320, 0){\includegraphics[scale=0.45]{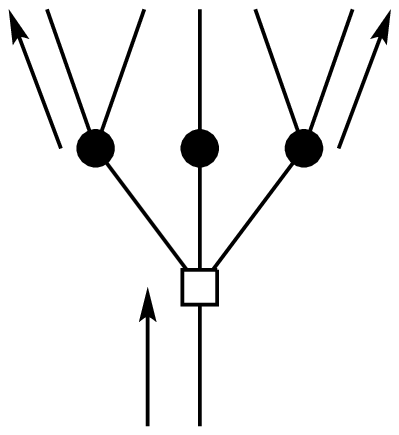}}
\put(50,20){\makebox(0,0){$(p, \bar{p})^T$}}
\put(80,160){\makebox(0,0){$\Ctree(i)(p, \bar{p})^T$}}
\put(210,20){\makebox(0,0){$(p, \bar{p})^T$}}
\put(240,160){\makebox(0,0){$\Vtree(i)(p, \bar{p})^T$}}
\put(370,20){\makebox(0,0){$(p, \bar{p})^T$}}
\put(400,200){\makebox(0,0){$\Vtree(i+1)\Ctree(i)(p, \bar{p})^T$}}
\end{picture}
\caption{\label{fig:CVtree}
Number of outgoing erased messages as a function of the probability of erasure
of the incoming message.
}
\end{figure}
The contribution of the edges in $\tree_1$ to $S$ is obtained by writing
\begin{align}
& \lim_{n \rightarrow \infty} \expectation[\mu_{1}^{(\ell)}\sum_{i\in\tree_1} \mu_{i}^{(\ell)}] \nonumber \\
= &  
\lim_{n \rightarrow \infty} \prob\{\mu_{1}^{(\ell)}=1\}\expectation[\sum_{i\in\tree_1} \mu_{i}^{(\ell)}\mid\mu_{1}^{(\ell)}=1]\label{equ:conditionedexpr}.
\end{align}
The conditional expectation on the right hand side is given by
\begin{align} 
1 + 
\sum_{j=1}^{\ell} (1,0)\Vtree(\ell)  \cdots \Ctree(\ell-j)\binom{1}{0}.\label{equ:conditioned}
\end{align}
where the $1$ is due to the fact that
$\expectation[\mu_{1}^{(\ell)}\mid\mu_{1}^{(\ell)}=1]=1$, and
each  summand
$(1, 0) \Vtree(\ell)  \cdots\Ctree(\ell-j)(1, 0)^T$,
is the expected number of erased messages in the $j$-{th} layer of edges 
in $\tree_1$, conditioned on the fact that the root edge is erased
at iteration $\ell$ (notice that $\mu_1^{(\ell)}=1$ implies 
$\mu_1^{(i)}=1$ for all $i\le \ell$).
Now multiplying (\ref{equ:conditioned}) by $\prob\{\mu_{1}^{(\ell)}=1\}=
x_{\ell}$ gives us (\ref{equ:tree1}).

The computation is similar for the edges in $\tree_3$ and results in
\begin{align*}
& \lim_{n \rightarrow \infty} \expectation[\mu_{1}^{(\ell)}\sum_{i\in\tree_3} \mu_{i}^{(\ell)}] = 
& x_{\ell} \sum_{j=1}^{2\ell} (1, 0)\Vtree(\ell)  \cdots  
\Ctree(\ell-j)\binom{1}{0}. 
\end{align*}
In this sum, when $j>\ell$, we have to evaluate the matrices 
$\Vtree(i)$ and $\Ctree(i)$ for negative
indices using the definitions given in (\ref{equ:Vtreeneg}) and
(\ref{equ:Ctreeneg}). The meaning of this case is simple: if $j > \ell$
then the observations in these layers do not influence the message
$\mu_{1}^{(\ell)}$. Therefore, for these steps we only need to {\em count}
the expected number of edges.  

In order to obtain $S$, it remains to compute the contribution of the edges
in $\tree_4$. This case is slightly more involved than the previous ones.
Recall that 
$\tree_4$ includes all the edges that are below the root node and point 
downwards. In Fig.~\ref{fig:tree}, edges $(e)$ and $(g)$ are 
elements of $\tree_4$. 
We claim that
\begin{align}
& \lim_{n \rightarrow \infty} \expectation[\mu_{1}^{(\ell)}\sum_{i\in\tree_4}
\mu_{i}^{(\ell)}] \nonumber \\
= &
 (1,0) \sum_{j=0}^{\ell}
\left(y_{\ell-j}U^{\star}(j, j) + 
(1-y_{\ell-j}) U^{0}(j, j) \right) \label{equ:tree4}\\ 
& +(1, 0)\sum_{j=\ell+1}^{2 \ell}
\Vtree(\ell) 
\cdots 
\Ctree(2 \ell-j)  \nonumber \\
& \quad \quad \quad \quad \quad \quad \cdot
\left(y_{\ell-j}U^{\star}(j, \ell) + 
(1-y_{\ell-j}) U^{0}(j, \ell) \right).\nonumber
\end{align}
The first sum on the right hand side
corresponds to messages $\mu_i^{(\ell)}$,  $i\in\tree_4$ 
whose computation tree contains the root variable node. 
In the case of 
Fig.~\ref{fig:tree}, where $\ell=2$, the contribution of edge $(e)$, would be 
counted in this first sum. The second term in (\ref{equ:tree4}) corresponds to 
edges $i\in\tree_4$, that are separated from the root edge by more than 
$\ell+1$ variable nodes. In Fig.~\ref{fig:tree}, edge $(g)$ is of this type.

In order to understand the first sum in (\ref{equ:tree4}), consider the root
edge and an edge $i\in\tree_4$ separated from the root edge by 
$j+1$ variable node with $j\in\{0,\cdots,\ell\}$. For this edge in $\tree_4$,
consider two messages it carries: the message that is sent from the 
variable node to the check node at the $\ell$-th iteration (this `outgoing'
message  participates in our second moment calculation), and the one 
sent from the check node to the variable node at the 
$(\ell-j)$-th iteration (`incoming').
Define the two-components vector $U^{\star}(j,j)$ as follows. Its first 
component is the joint probability that both the root and the 
outgoing messages 
are erased conditioned on the fact that the incoming message is erased, 
multiplied by the expected
number of edges in $\tree_4$ whose distance from the root 
is the same as for edge $i$.
Its second component is the joint probability that the root message is erased 
and that the outgoing message is known, again conditioned on  the 
incoming message being erased, and multiplied by the expected
number of edges in $\tree_4$ at the same distance from the root.
The vector $U^0(j,j)$ is defined in exactly the same manner except that 
in this case we condition on the incoming message  being
\emph{known}. 
The superscript $\star$ or $0$ accounts respectively for the cases 
where the incoming message is erased or known. 

From these definitions, it is clear 
that the contribution to $S$ of the edges that are in $\tree_4$ and separated 
from the root edge by $j+1$ variable nodes with $j\in\{0,\cdots,\ell\}$, 
is $(1,0)\left(y_{\ell-j}U^{\star}(j, j) + (1-y_{\ell-j}) U^{0}(j, j) \right)$.
We still have to evaluate $U^{\star}(j,j)$ and $U^{0}(j,j)$. 
In order to do this, we define the vectors 
$U^{\star}(j,k)$ and $U^{0}(j,k)$ with $k\leq j$, analogously to the case 
$k=j$, except that this time we consider the root edge and an edge
in $i\in\tree_4$ separated from the root edge by $k+1$ variable nodes. 
The outgoing
message we consider is the one at the $(\ell-j+k)$-th iteration and the 
incoming message we condition on, is the one at the $(\ell-k)$-th iteration.
It is easy to check that $U^{\star}(j,j)$ and $U^0(j,j)$ can be computed in a 
recursive manner using $U^{\star}(j,k)$ and $U^0(j,k)$.
The initial conditions are 
\begin{align*}
U^{\star}(j,0) = & \binom{y_{\ell-j} \epsilon \ledge'(y_{\ell})}{(1-y_{\ell-j})  \epsilon \ledge'(y_{\ell})}, \quad U^{0}(j,0) = & \binom{0}{0},
\end{align*}
and the recursion is for $k\in\{1,\cdots,j\}$ is the one given in 
Lemma~\ref{lem:varianceforfinitel}, cf. Eqs.~(\ref{eq:RecT4_1})
and (\ref{eq:RecT4_2}).
Notice that any received value which is on the path between the root 
edge and the edge in $\tree_4$ affects both the messages $\mu_1^{(\ell)}$
and $\mu_i^{(\ell)}$ on the corresponding edges.
This is why this recursion is slightly more involved than the one  
for $\tree_1$. 
The situation is depicted in the left side of Fig.~\ref{fig:coverlap}.
\begin{figure}[htp]
\centering
\setlength{\unitlength}{0.22bp}%
\begin{picture}(800, 600)
\put(0, 0){\includegraphics[scale=0.22]{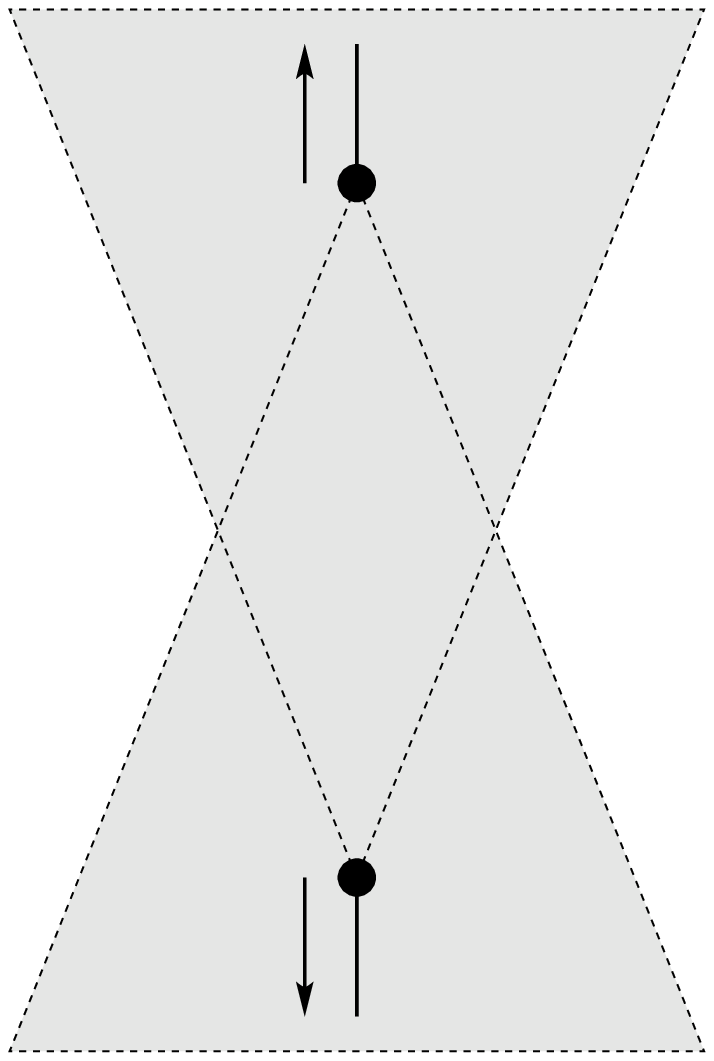}}
\put(500, 0){\includegraphics[scale=0.22]{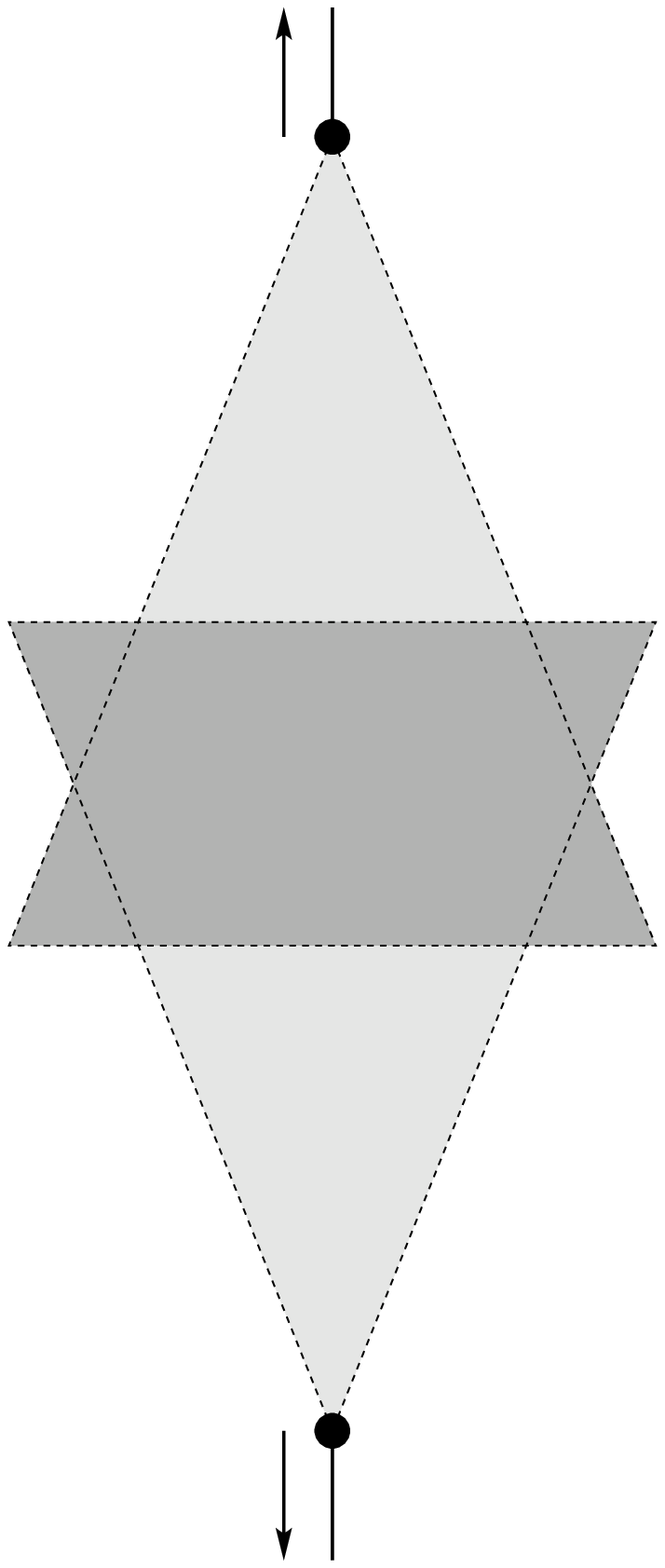}}
\put(35,400){\makebox(0,0){root edge}}
\put(535,500){\makebox(0,0){root edge}}
\put(35,200){\makebox(0,0){edge in $\tree_4$}}
\put(535,100){\makebox(0,0){edge in $\tree_4$}}
\end{picture}
\caption{\label{fig:coverlap} The two situations that arise when computing
the contribution of $\tree_4$. 
In the left side we show the case where the two edges are separated by at most $\ell+1$ 
variable nodes and in the right side, the case where they are separated by
more than $\ell+1$ variable nodes.}
\end{figure}

Consider now the case of edges in $\tree_4$ that are separated from 
the root edge by more than $\ell+1$ variable nodes, cf. right picture in 
Fig.~\ref{fig:coverlap}. In this case, not all of the received values along
the  path connecting the two edges, do affect both messages. 
We therefore have to adapt the
previous recursion. We start from the root edge and compute the effect of 
the received values that only affect this message resulting in a expression 
similar to the one we used to compute the contribution of $\tree_1$. This 
gives us the following initial condition
\begin{align*}
U^{\star}(j,j-\ell) = & 
(1, 0) 
\Vtree(\ell) 
\cdots 
\Ctree(2\ell-j) (1, 0)^T \binom{\epsilon\ledge'(y_{2\ell-j})}{0}\\
& \!\!\!\!\!\!\!\!\!\!\!\!\!\!\!\!\!\!\!\!\!\!\!\!\!\!\!\!+ (1, 0) \Vtree(\ell) 
\cdots 
\Ctree(2\ell-j) (0, 1)^T \binom{\epsilon(\lambda'(1)-\lambda'(y_{2\ell-j}))}{
\lambda'(1)(1-\epsilon)} ,\\
U^{0}(j,j-\ell) = &  (1, 0) \Vtree(\ell) 
\cdots 
\Ctree(2\ell-j) (0, 1)^T
\binom{\epsilon\lambda'(1)}{(1-\epsilon)\lambda'(1)} .
\end{align*}
We then apply the recursion given in Lemma~\ref{lem:varianceforfinitel} to the 
intersection of the computation trees. We have to stop the recursion at 
$k=\ell$ (end of the intersection of the computation trees). It remains to 
account for the received values that only affect the messages on the edge in 
$\tree_4$. This is done by writing
\begin{align*}
(1, 0)\sum_{j=\ell+1}^{2 \ell}
\Vtree(\ell)  \cdots &\Ctree(2 \ell-j) \\
&\cdot \left(y_{\ell-j}U^{\star}(j, \ell) + 
(1-y_{\ell-j}) U^{0}(j, \ell) \right),
\end{align*}
which is the second term on the right hand side of Eq.~(\ref{equ:tree4}).

\subsubsection{Computation of $S^c$}
We still need to compute
$S^c=\lim_{n \rightarrow \infty} \left(\expectation[\mu_{1}^{(\ell)}\sum_{i\in\btree} \mu_{i}^{(\ell)}]  - n \lnodenormalized'(1) x_{\ell}^2  \right)$.
Recall that by definition, all the messages that are carried by edges 
in $\tree^c$ at the $\ell$-th iteration are functions of a set of
received values distinct  from the ones $\mu_1^{(\ell)}$ depends on. 
At first sight, one might think that such messages are independent from 
$\mu_1^{(\ell)}$. This is indeed the case when the Tanner graph is 
regular, i.e. for the degree distributions $\ledge(x)=x^{\dl-1}$
and $\redge(x)=x^{\dr-1}$. We then have 
\begin{align*}
S^c = &\lim_{n \rightarrow \infty} \left(\expectation[\mu_{1}^{(\ell)}\sum_{i\in\btree} \mu_{i}^{(\ell)}]  - n \lnodenormalized'(1) x_{\ell}^2  \right)\\
= & \lim_{n \rightarrow \infty} \left( |\btree|x_{\ell}^2-\lnode'(1)x_{\ell}^2\right)\\
= & \lim_{n \rightarrow \infty} \left( (\lnode'(1)-|\tree|)x_{\ell}^2-\lnode'(1)x_{\ell}^2\right)\\
= & -|\tree|x_{\ell}^2
\end{align*}
with the cardinality of $\tree$ being $|\tree|=\sum_{i=0}^{2\ell}{(\dl-1)^i(\dr-1)^i \dl}$ $+\sum_{i=1}^{\ell}{(\dl-1)^{i-1}(\dr-1)^{i} \dl}$.

Consider now an irregular ensemble and let $\grapht$ be the graph composed 
by the edges in $\tree$ and by the variable 
and check nodes connecting them.
Unlike in the regular case,  $\grapht$ is not fixed anymore and depends 
(in its size as well as in its structure) on the graph realization. 
It is clear that the root message $\mu_1^{(\ell)}$ 
depends on the realization of $\grapht$. We will see that the 
messages carried by the edges in $\btree$ also depend on the realization of
$\grapht$. On the other hand they are clearly conditionally independent
given $\grapht$ (because, conditioned on $\grapht$, $\mu_1^{(\ell)}$ 
is just a deterministic function of the received symbols in its computation 
tree). If we let $j$ denote a generic edge in $\btree$ (for instance, the one
with the lowest index), we can therefore write
\begin{align}
S^c = &\lim_{n \rightarrow \infty} \left(\expectation[\mu_{1}^{(\ell)}\sum_{i\in\btree} \mu_{i}^{(\ell)}]  - n \lnodenormalized'(1) x_{\ell}^2  \right)\nonumber\\
= &\lim_{n \rightarrow \infty} \left(\expectation_{\grapht}[\expectation[\mu_{1}^{(\ell)}\sum_{i\in\btree} \mu_{i}^{(\ell)}\mid \grapht]]  - n \lnodenormalized'(1) x_{\ell}^2  \right)\nonumber\\
= &\lim_{n \rightarrow \infty} \left(\expectation_{\grapht}[|\btree|\expectation[\mu_{1}^{(\ell)}\mid \grapht]  \expectation[\mu_{j}^{(\ell)}\mid \grapht]]  - n \lnodenormalized'(1) x_{\ell}^2  \right) \nonumber\\
= &\lim_{n \rightarrow \infty} \left(\expectation_{\grapht}[ \left(n\lnodenormalized'(1)-|\tree|\right) \expectation[\mu_{1}^{(\ell)}\mid \grapht] \expectation[\mu_{j}^{(\ell)}\mid \grapht]] \right.\nonumber \\
&\left.\phantom{\expectation[\mu_{1}^{(\ell)}\mid \grapht]}\quad \quad \quad\quad \quad \quad\quad \quad \quad \quad \quad \quad 
- n \lnodenormalized'(1) x_{\ell}^2  \right)\nonumber\\
= &\lim_{n \rightarrow \infty} n\lnodenormalized'(1) \left(\expectation_{\grapht}[\expectation[\mu_{1}^{(\ell)}\!\mid \grapht] \expectation[\mu_{j}^{(\ell)}\!\mid \grapht]]  - n \lnodenormalized'(1) x_{\ell}^2  \right)\nonumber\\
& - \lim_{n \rightarrow \infty} \expectation_{\grapht}[ |\tree| \expectation[\mu_{1}^{(\ell)}\mid \grapht] \expectation[\mu_{j}^{(\ell)}\mid \grapht]].
\label{equ:sizeeffect}
\end{align}
We need to compute $\expectation[\mu_{j}^{(\ell)}\mid \grapht]$
for a fixed realization of $\grapht$ and an arbitrary edge $j$ taken from 
$\btree$ (the expectation does not depend on $j\in\btree$: we can therefore
consider it as a random edge as well). 
This value differs slightly from $x_{\ell}$ for two reasons.
The first one is that we are 
dealing with a fixed-size Tanner graph (although taking later the limit 
$n\rightarrow \infty$) and therefore the degrees of the nodes in 
$\grapht$ are correlated with the degrees of nodes in its complement 
$\graph\backslash\grapht$. Intuitively, if $\grapht$ contains
an unusually large number of high degree variable nodes, the rest of the graph
will contain an unusually small number of high degree variable nodes affecting
the average $\expectation[\mu_{j}^{(\ell)}\mid \grapht]$. 
 The second reason why
$\expectation[\mu_{j}^{(\ell)}\mid \grapht]$ differs from $x_{\ell}$, is that 
certain messages carried by edges in $\btree$ which are close to $\grapht$ are
affected by messages that flow out of $\grapht$. 

The first effect can
be characterized by computing the degree distribution on 
$\graph\backslash\grapht$ as a function of $\grapht$. Define 
$V_i(\grapht)$ (respectively $C_i(\grapht)$) to be the number of variable 
nodes (check nodes) of degree $i$ in $\grapht$, and let
$V(x;\grapht)=\sum_i V_i(\grapht) x^i$ and $C(x;\grapht)=\sum_i C_i(\grapht) 
x^i$. We shall also need the derivatives of 
these polynomials:
$V'(x;\grapht)=\sum_i i V_i(\grapht) x^{i-1}$ and 
$C'(x;\grapht)=\sum_i i C_i(\grapht) x^{i-1}$.
It is easy to check that if we take a bipartite graph having a variable degree 
distributions $\ledge(x)$ and remove a variable node of degree $i$, the 
variable degree distribution changes by
\begin{align*}
\delta_i \ledge(x) & = \frac{i \ledge(x) -i x^{i-1}}{n \lnodenormalized'(1)} + O(1/n^2).
\end{align*}
Therefore, if we remove $\grapht$ from the bipartite graph, the remaining graph
will have a variable perspective degree distribution that differ from the 
original by
\begin{align*}
\delta \ledge(x) & = \frac{V'(1;\grapht) \ledge(x) - V'(x;\grapht)}{n \lnodenormalized'(1)} + O(1/n^2).
\end{align*}
In the same way, the check degree distribution when we remove $\grapht$ changes
by
\begin{align*}
\delta \redge(x) & = \frac{C'(1;\grapht) \redge(x) - C'(x;\grapht)}{n \lnodenormalized'(1)} + O(1/n^2).
\end{align*}
If the degree distributions
change  by $\delta \ledge(x)$ and $\delta \redge(x)$, the fraction $x_{\ell}$ 
of erased variable-to-check messages changes by $\delta x_{\ell}$.
To the linear order we get
\begin{align*}
\delta x_{\ell} = & 
\sum_{i=1}^{\ell}
\prod_{k=i+1}^{\ell}\epsilon \ledge'(y_{k}) \redge'(\bar{x}_{k-1})
\left[\epsilon \delta \ledge(y_i)
- \epsilon \ledge'(y_i) \delta \redge(\bar{x}_{i-1}) \right],\\
 = & 
\frac{1}{\lnode'(1)}\sum_{i=1}^{\ell}
F_i\left[\epsilon (V'(1;\grapht) \ledge(y_i) - V'(y_i;\grapht)) \right.\\
& \!\!\!\!\!\left. - \epsilon \ledge'(y_i) (C'(1;\grapht) \redge(\bar{x}_{i-1}) - 
C'(\bar{x}_{i-1};\grapht))
\right] + O(1/n^2),
\end{align*}
with $F_i$ defined as in Eq.~(\ref{equ:Fi}).

Imagine now that we ix the degree distribution 
of $\graph\backslash \grapht$. The conditional expectation  
$\expectation[\mu_{j}^{(\ell)}\mid \grapht]$ still 
depends on the detailed structure of $\grapht$.
The reason is that the messages that flow out of the boundary of 
$\grapht$ (both their number and value) depend on $\grapht$, and these
message affect messages in $\graph\backslash \grapht$. 
Since the fraction of such (boundary) messages is $O(1/n)$, their effect can be 
evaluated again perturbatively.

Call ${\cal B}$ the number of edges forming the boundary of $\grapht$ 
(edges emanating upwards from the variable nodes that are 
$\ell$ levels above the root edge and emanating downwards 
from the variable nodes that are $2\ell$ levels 
below the root variable node) and let ${\cal B}^{\star}_i$ be the number of 
erased messages carried at the $i$-th iteration by these edges. 
Let $\tilde{x}_i$ be the fraction of erased messages, incoming to check nodes 
in  $\graph\backslash\grapht$ from variable nodes in $\graph\backslash\grapht$,
at the $i$-th iteration. Taking into account the messages coming from 
variable nodes in $\grapht$ (i.e. corresponding to boundary edge),
the overall fraction will be $\tilde{x}_i+\delta\tilde{x}_i$, where
\begin{align*}
\delta\tilde{x}_i =
\frac{ {\cal B}^{\star}_i-  {\cal B} \tilde{x}_i}
{n\lnodenormalized'(1)} + O(1/n^2).
\end{align*}
This expression simply comes from the fact that at the $i$-th iteration,
we have $(n\lnode'(1)-\tree) = n\lnode'(1)(1+O(1/n))$ messages in 
the complement of $\grapht$ of which a fraction $\tilde{x}_i$ is erased. 
Further ${\cal B}$ messages incoming from the boundaries of which 
 ${\cal B}^{\star}_i$ are erasures.

Combining the two above effects, we have
 for an edge $j\in \btree$ 
\begin{align*}
& \expectation[\mu_{j}^{(\ell)}\mid \grapht] = 
x_{\ell} +  \frac{1}{\lnode'(1)}
\sum_{i=1}^{\ell}F_i
\left[
x_i V'(1;\grapht) - \epsilon V'(y_i;\grapht)  \right. \\
& \left. \quad \quad \quad \quad \quad -\epsilon \ledge'(y_i)  (C'(1;\grapht) \redge(\bar{x}_{i-1}) - 
C'(\bar{x}_{i-1};\grapht))
\right]\\
& \quad \quad \quad \quad \quad+ \frac{1}{\lnode'(1)} \sum_{i=1}^{\ell-1} F_i \left({\cal B}^{\star}_i-  {\cal B} x_i \right) + O(1/n^2).
\end{align*}
We can now use this expression  (\ref{equ:sizeeffect}) to obtain
\begin{align*}
S^c &= \lim_{n \rightarrow \infty} \lnode'(1) 
\left(
\expectation_{\grapht}[\expectation[\mu_{1}^{(\ell)}\mid \grapht] \expectation[\mu_{j}^{(\ell)}\mid \grapht]]  - n \lnodenormalized'(1) x_{\ell}^2  
\right)\\
& - \lim_{n \rightarrow \infty} \expectation_{\grapht}[ |\tree| \expectation[\mu_{1}^{(\ell)}\mid \grapht] \expectation[\mu_{j}^{(\ell)}\mid \grapht]]\\
&= \sum_{i=1}^{\ell}F_i \left( x_i   \expectation[\mu_{1}^{(\ell)} V'(1;\grapht)] - \epsilon  \expectation[\mu_{1}^{(\ell)} V'(y_i;\grapht)]\right)\\
&  - \sum_{i=1}^{\ell}F_i \epsilon \ledge'(y_i) \left( \expectation[\mu_{1}^{(\ell)} C'(1;\grapht)] \redge(\bar{x}_{i-1}) \right. \\
& \left.\quad \quad \quad \quad \quad \quad \quad \quad \quad \quad \quad \quad -\expectation[\mu_{1}^{(\ell)}  C'(\bar{x}_{i-1};\grapht)] \right)\\
& + \sum_{i=1}^{\ell-1} F_i  \expectation[\mu_{1}^{(\ell)} {\cal B}^{\star}_i]
 - \sum_{i=1}^{\ell-1} F_i  x_i \expectation[\mu_{1}^{(\ell)}{\cal B}]
 - x_{\ell} \expectation[\mu_{1}^{(\ell)} \Vgtp(1)],
\end{align*}
where we took the limit $n\rightarrow \infty$ and replaced 
$|\tree|$ by $V'(1;\grapht)$.

It is clear what each of these values represent. For example, $\expectation[\mu_{1}^{(\ell)} V'(1;\grapht)]$ is the expectation of $\mu_1^{(\ell)}$ times
the number of edges that are in $\grapht$. Each of these terms can be computed 
through recursions that are similar in spirit to the ones used to compute $S$. 
These recursions are provided in the body of 
Lemma~\ref{lem:varianceforfinitel}.
We will just explain in further detail how the terms 
$\expectation[\mu_{1}^{(\ell)}{\cal B}]$ and $\expectation[\mu_{1}^{(\ell)} {\cal B}^{\star}_i]$ are computed. 

We claim that 
\begin{align*}
\expectation[\mu_{1}^{(\ell)}{\cal B}] = & \left(x_{\ell}+
(1, 0)\Vtree(\ell)  \cdots \Ctree(0)\Vtree(0)(1,0)^T\right) \left(\ledge'(1)\redge'(1)\right)^{\ell}.
\end{align*}
The reason is that $\mu_1^{(\ell)}$ depends only on the realization
of its computation tree and not on the whole $\grapht$. 
From the definitions of $\grapht$,
the boundary of $\grapht$ is in average $(\ledge'(1)\redge'(1))^{\ell}$ larger
than the boundary of the computation tree. 
Finally, the
expectation of $\mu_1^{(\ell)}$ times  the number of edges in the 
boundary of its computation tree is 
computed analogously to what has been done for the contribution of $S$.
The result is
$\left(x_{\ell}+
(1, 0)\Vtree(\ell)  \cdots \Ctree(0)\Vtree(0)(1,0)^T\right)$
(the term $x_{\ell}$ accounts for the root edge, and the other one of the 
lower boundary of the computation tree). 
Multiplying this by $(\ledge'(1)\redge'(1))^{\ell}$, we obtain the
above expression.

The calculation of $\expectation[\mu_{1}^{(\ell)} {\cal B}^{\star}_i]$ is 
similar. We start by computing the expectation of $\mu_1^{(\ell)}$ 
multiplied by the number of edges in the boundary of its computation tree. This
number has to be multiplied by $(1, 0) \Vtree(i) \Ctree(i-1) \cdots \Vtree(i-\ell+1) \Ctree(i-\ell) (1, 0)^T$ to account for what happens between the boundary
of the computation tree and the boundary of $\grapht$. We  therefore obtain
\begin{align*}
\expectation[\mu_{1}^{(\ell)}{\cal B}^{\star}_i] = & \left(x_{\ell}+
(1, 0)\Vtree(\ell) 
\cdots 
\Ctree(0)\Vtree(0)(1,0)^T\right)\\
& \cdot (1, 0) \Vtree(i) 
\cdots 
\Ctree(i-\ell) (1, 0)^T.
\end{align*}
\end{proof}

The expression provided in the above lemma has been used to plot 
$\variance^{(\ell)}$ for $\epsilon \in (0,1)$ and for several values of $\ell$ 
in the case of an irregular ensemble in Fig.~\ref{fig:varianceforfinitel}.

It remains to determine the asymptotic behavior of this quantity as
the number of iterations converges to infinity. 
\blemma
Let $\graph$  be chosen uniformly at random from $\eldpc n \ledge \redge$ and consider
transmission over the BEC of erasure probability $\epsilon$. Label the 
$n \lnodenormalized'(1)$ edges of
$\graph$ in some fixed order by the elements of $\{1,\dots ,
n \lnodenormalized'(1)\}$. 
Set $\mu^{(\ell)}_i$ equal to one if the  message along edge 
$i$ from  variable to check node, after $\ell$ iterations,  is an erasure
and equal to zero otherwise.
Then
\begin{align*}
&
\lim_{\ell \rightarrow \infty}
\lim_{n \rightarrow \infty}
\frac{\expectation[(\sum_{i} \mu_{i}^{(\ell)})^2 ]-
\expectation[(\sum_{i} \mu_{i}^{(\ell)})]^2}{n \lnodenormalized'(1)} =  \\
& + \frac{\epsilon^2 \ledge'(y)^2(
\redge(\bar{x})^2-\redge(\bar{x}^2)+
\redge'(\bar{x})(1-2 x \redge(\bar{x}))-
\bar{x}^2 \redge'(\bar{x}^2))
}{
(1- \epsilon \ledge'(y) \redge'(\bar{x}))^2} 
\\
&
+ \frac{\epsilon^2 \ledge'(y)^2 \redge'(\bar{x})^2 (
\epsilon^2 \ledge(y)^2-\epsilon^2 \ledge(y^2) - y^2 \epsilon^2 \ledge'(y^2))}{
(1- \epsilon \ledge'(y) \redge'(\bar{x}))^2}  \\
& + \frac{(x-\epsilon^2 \ledge(y^2)-y^2 \epsilon^2 \ledge'(y^2))(1+\epsilon \ledge'(y) \redge'(\bar{x}))+\epsilon y^2 \ledge'(y) }{1- \epsilon \ledge'(y) \redge'(\bar{x})}. 
\end{align*}
\end{lemma}

The proof is a (particularly tedious) calculus exercise, and we omit it here
for the sake of space.
%

\bibliographystyle{unsrt}

\begin{thebibliography}{1}

\bibitem{RiU05}
T.~Richardson and R.~Urbanke.
\newblock {\em Modern Coding Theory}.
\newblock Cambridge University Press, 2006.
\newblock In preparation.

\bibitem{Mon01b}
A.~Montanari.
\newblock Finite-size scaling of good codes.
\newblock In {\em Proc. 39th Annual Allerton Conference on Communication,
  Control and Computing}, Monticello, IL, 2001.

\bibitem{AMRU04}
A.~Amraoui, A.~Montanari, T.~Richardson, and R.~Urbanke.
\newblock Finite-length scaling for iteratively decoded ldpc ensembles.
\newblock submitted to IEEE IT, June 2004.

\bibitem{AMRU04-allerton}
A.~Amraoui, A.~Montanari, T.~Richardson, and R.~Urbanke.
\newblock Finite-length scaling and finite-length shift for low-density
  parity-check codes.
\newblock In {\em Proc. 42th Annual Allerton Conference on Communication,
  Control and Computing}, Monticello, IL, 2004.

\bibitem{AMU05}
A.~Amraoui, A.~Montanari, and R.~Urbanke.
\newblock Finite-length scaling of irregular {LDPC} code ensembles.
\newblock In {\em Proc. IEEE Information Theory Workshop}, Rotorua,
  New-Zealand, Aug-Sept 2005.

\bibitem{LMSSS97}
M.~Luby, M.~Mitzenmacher, A.~Shokrollahi, D.~A. Spielman, and V.~Stemann.
\newblock Practical loss-resilient codes.
\newblock In {\em Proceedings of the $29$th annual ACM Symposium on Theory of
  Computing}, pages 150--159, 1997.

\bibitem{MMU05}
C.~M{\'e}asson, A.~Montanari, and R.~Urbanke.
\newblock Maxwell's construction: The hidden bridge between maximum-likelihood
  and iterative decoding.
\newblock submitted to IEEE Transactions on Information Theory, 2005.

\end{thebibliography}

\newcommand{\SortNoop}[1]{}

\end{document}